\documentclass{aa}  
\usepackage{graphicx}
\usepackage{txfonts}
\usepackage{float}
\usepackage{lscape}
\usepackage{color, soul}
\usepackage{lineno}

\soulregister\cite7
\soulregister\ref7
\soulregister\pageref7
\usepackage{natbib}

\begin{document}

   \title{Reconstruction of spectral solar irradiance since 1700 from simulated magnetograms}

   \author{M. Dasi-Espuig\inst{1,2},
          J. Jiang\inst{3},
	  N. A. Krivova\inst{2},
          S. K. Solanki\inst{2,4},
          Y. C. Unruh\inst{1}
          \and
	  K. L. Yeo\inst{2}
          }

   \offprints{}

   \institute{Astrophysics Group, Blackett Laboratory, Imperial College London, SW7 2AZ, UK\\
                  \email{m.dasi-espuig@imperial.ac.uk}
          \and
             {Max-Planck-Institut f\"ur Sonnensystemforschung,
                   Justus-von-Liebig-Weg 3, 37077 G\"ottingen, Germany\\
                   }
         \and
              {Key Laboratory of Solar Activity, National Astronomical Observatories, Chinese Academy of Sciences, Beijing 100012, China\\
             }
         \and
              School of Space Research, Kyung Hee University, Yongin, Gyeonggi, 446-701, Korea\\
             }
 
   \date{Received ; accepted }

 
  \abstract
   {}
   {We present a reconstruction of the spectral solar irradiance since 1700 using the SATIRE-T2 (Spectral And Total Irradiance REconstructions for the Telescope era version 2) model. This model uses as input magnetograms simulated with a surface flux transport model fed with semi-synthetic records of emerging sunspot groups.}
   {The record of sunspot group areas and positions from the Royal Greenwich Observatory (RGO) is only available since 1874. We used statistical relationships between the properties of sunspot group emergence, such as the latitude, area, and tilt angle, and the sunspot cycle strength and phase to produce semi-synthetic sunspot group records starting in the year 1700.
The semi-synthetic records are fed into a surface flux transport model to obtain daily simulated magnetograms that map the distribution of the magnetic flux in active regions (sunspots and faculae) and their decay products on the solar surface. The magnetic flux emerging in ephemeral regions is accounted for separately based on the concept of extended cycles whose length and amplitude are linked to those of the sunspot cycles through the sunspot number.
The magnetic flux in each surface component (sunspots, faculae and network, and ephemeral regions) was used to compute the spectral and total solar irradiance between the years 1700 and 2009.
This reconstruction is aimed at timescales of months or longer although the model returns daily values.}
   {We found that SATIRE-T2, besides reproducing other relevant observations such as the total magnetic flux, reconstructs the total solar irradiance (TSI) on timescales of months or longer in good agreement with the PMOD composite of observations, as well as with the reconstruction starting in 1878 based on the RGO-SOON data. The model predicts an increase in the TSI of 1.2$^{+0.2}_{-0.3}$ Wm$^{-2}$ between 1700 and the present. The spectral irradiance reconstruction is in good agreement with the UARS/SUSIM measurements as well as the Lyman-$\alpha$ composite. 
 }
   {}

   \keywords{sunspots,  surface magnetism, solar cycle, solar-terrestrial relations
               }

   \authorrunning{Dasi-Espuig et al.}
   \titlerunning{Spectral Solar Irradiance since 1700}

   \maketitle
%

\section{Introduction}

Solar irradiance is the main external energy source of the Earth's climate system and, therefore, measurements of its variation are crucial to disentangle the influence of the Sun on climate from other sources.
The available measurements \citep[for recent reviews see][]{ermolli13, kopp14, yeo14-b} are accurate enough to clearly show a relative change in the total solar irradiance (TSI) of up to $\sim$0.3\% and $\sim$0.1\% on timescales of the solar rotation and the 11-year solar cycle, respectively. This relative variation is wavelength-dependent, increasing strongly towards shorter wavelengths.
Although wavelengths above 400 nm (i.e. in the visible and infrared) contribute more than 90\% to the total solar irradiance, the relative variability at these spectral ranges over the solar cycle is only of the order of 0.1\%.
In contrast, between 200 and 300 nm the relative variability over the solar cycle is 1 -- 10\%, and reaches 50\% or more at wavelengths below 200 nm \citep{floyd03, krivova06}.

While direct measurements of the total and spectral solar irradiance cover the past four decades almost continuously, climate studies require even longer time series to investigate, for example, the effects of irradiance changes in the Ocean-Atmosphere coupling \citep[see][and references therein]{haigh01, solanki13}.
Additionally, instrumental degradation, sensitivity changes and differences in the absolute values of the various radiometers introduce uncertainties in the observed trends of the irradiance variation on timescales longer than the solar cycle \citep[e.g.][]{ermolli13}. 
In such cases when longer and continuous time series of irradiance are needed, models are crucial to extend irradiance records into the past.

Most successful models of solar irradiance assume that its variation on timescales longer than a day is due to the evolution of the magnetic fields on the solar surface \citep[e.g.][]{foukal88, fligge00, krivova03, wang05, crouch08}. The SATIRE (Spectral and Total Irradiance REconstructions) set of models \citep{krivova11-b} are based on this assumption. The SATIRE-S (S for Satellite era) model in particular supports this idea since it is based on direct and detailed 
measurements of the solar surface magnetic fields (magnetograms) and has been able to reproduce more than 90\% of the measured variation of total solar irradiance \citep{krivova03, wenzler06, ball12, yeo14}.
The recording of daily full-disc magnetograms started in 1974, and therefore, proxies of the photospheric magnetic fields are needed to extend the models further into the past.

SATIRE-T \citep[for Telescope era;][]{krivova07, krivova10} employs a physical model to calculate the time-varying photospheric magnetic flux integrated over the whole solar surface, from the sunspot 
number \citep{solanki02-b, vieira10}. This model can thus reconstruct irradiance back to the Maunder Minimum.
The next step towards realism was taken by \cite{Dasi-Espuig14}, who presented the SATIRE-T2 model (for the Telescope era, version 2) that uses magnetograms simulated with a surface flux transport model based on sunspot group areas and positions \citep{cameron10, jiang10, jiang11}. This model allowed a reconstruction of the TSI starting in 1878. Employing the simulated magnetograms improved the agreement between the modeled and measured TSI on rotational timescales, since the surface flux transport model uses the information on the spatial distribution of the magnetic features and thus allows a more realistic representation of the evolution of magnetic fields on the solar surface. It also accounts for the centre-to-limb variation of the contrast of magnetic features.

Here we present an extension of the SATIRE-T2 model to reconstruct the total and spectral solar irradiance between 115nm and 160$\mu$m starting in 1700. 
Since the sunspot group area record starts in 1874, to extend the reconstructions back to 1700 we used the same surface flux transport model as in \cite{Dasi-Espuig14}, but this time fed with the semi-synthetic sunspot group records 
following \cite{jiang11-a, jiang11}. These records obey the statistical relationships between the properties of sunspot group emergence seen in the Royal Greenwich Observatory (RGO) dataset (such as the mean latitude, latitude distribution, areas, and tilt angles, e.g., \citealt{solanki08, Dasi-Espuig10, Dasi-Espuig13}), and the cycle strength and phase obtained from the sunspot group number.

The paper is structured as follows. The surface flux transport model and the semi-synthetic sunspot group records are introduced in Sect. 2. In Sect. 3 we briefly describe the SATIRE-T2 model, validate it against available observations of the total magnetic flux and TSI, and explain the extension going back to 1700. Section 4 presents the results of the total and spectral irradiance reconstructions since the end of the Maunder Minimum and compares them to previous studies. Finally, in Sect. 5 we summarise our results.


 \section{The photospheric magnetic flux}
 
 \subsection{Active region flux}
 
The surface flux transport model (SFTM) describes the transport of the large-scale magnetic fields in the photosphere under the effects of large-scale surface flows: differential rotation, meridional flow, and turbulent diffusivity \citep[for a review of the model see][]{jiang14}. The sources of magnetic flux in the SFTM are the active regions, which are composed of sunspots and faculae.
For sunspot areas and positions we used semi-synthetic records of sunspot groups derived from the sunspot group number \citep{hoyt98}, R$_{g}$, starting in the year 1700. The facular areas and positions were then computed based on the sunspot group areas and positions as described below.
The model parameters used here are taken from observations and are the same as in \cite{jiang11} and \cite{Dasi-Espuig14}.

The semi-synthetic sunspot group records are based on the statistical relationships between the properties of sunspot group emergence and cycle strength and phase derived by \cite{jiang11-a}. These records include area, position, and tilt angle of sunspot groups at the time when each group reaches its maximum area.
\cite{jiang11-a} used data from the Royal Greenwich Observatory (RGO), the sunspot number and the tilt angle data from Mount Wilson and Kodaikanal Observatories \citep[e.g][]{howard84, sivaraman93}.
From the RGO record the authors selected groups only once, at the time they reached the maximum area; from the sunspot number they obtained the strength and phase of a cycle; and from the tilt angle data they obtained the mean tilt angle of sunspot groups emerging in each cycle.
They then derived statistical relationships that describe the dependence of the mean latitude, the width of the latitude distribution, the mean tilt angle, and the area distribution of the observed sunspot groups at the time they reach maximum area, with the strength and phase of a cycle. This was done for the time when there is an overlap between the time series, i.e. 1874 to 1976 (cycles 12 -- 20) for the RGO data and 1913 to 1986 (cycles 15 -- 21) for the tilt angle data.
Finally, the total number of groups emerging in a month ($N_{SG}$) was normalised to match the number of groups emerging in a month as recorded in the RGO data set (N$_{SG}$ = R$_{g}$/2.1).
The sunspot groups were then distributed randomly in time throughout the days of a month.

We extrapolated the statistical relationships described above to extend the RGO record back to 1700. We thereby implicitly assume that the dynamo has operated in a similar way over the past three centuries. Note that during the Maunder minimum, i.e. before 1700, the dynamo might have operated in a different manner \citep{usoskin01, owens12, usoskin15-b}.

An individual semi-synthetic record containing the time of maximum area, latitude, area and tilt angle was drawn from a population that obeys the observed statistical relationships. The longitude positions were obtained by distributing groups randomly within $\pm70^{\circ}$ of the central meridian for each day. 
Although we used the same SFTM as in \cite{jiang11}, here we did not consider active longitudes when constructing the semi-synthetic records. Thus the open flux during cycle maxima is weaker than computed by \cite{jiang11} and reconstructed from geomagnetic data by \cite{lockwood09-c, lockwood14} that focused on the evolution of the open flux. The difference is within 20\% at these times for each cycle \citep[c.f. Fig. 6 of][]{jiang11}.
The active longitudes are an important ingredient in the evolution of the solar open flux during cycle maxima, but have no significant effect on the total magnetic flux \citep{wang02-b} and therefore we do not expect significant changes in irradiance either.

For each active region, the sunspot group area, $A_s$, is provided by the semi-synthetic records, and the facular area, $A_f$, was computed using the empirical relationship taken from \cite{chapman97} between sunspot and facular areas. We assume that the ratio of facular to sunspot areas has not changed from cycle to cycle over the past $\sim$3 centuries. The total area of the emerging active region, $A_R$, was thus obtained as $A_R = A_s + A_f$.

The SFTM produces a daily map of the magnetic field on the photosphere for the whole Sun (i.e. a cylindrical map) which we convert into simulated magnetograms by projecting the portion of the solar surface facing the Earth each day (within $\pm90^{\circ}$ of the central meridian) onto a disc \citep[for details see][]{Dasi-Espuig14}.

 \subsection{Ephemeral region flux}
 
The SFTM does not include magnetic flux emerging in ephemeral regions and therefore the ephemeral region magnetic flux ($\Phi_{ER}$) was modeled separately in the same manner as in \cite{Dasi-Espuig14} following \cite{krivova07, krivova10} and \cite{vieira10}.
Detailed studies of ephemeral regions in cycles 19 -- 23 have shown that the number of such regions varies cyclically over time in phase with the solar cycle, but their duration is longer, starting before and ending after the corresponding sunspot cycle \citep{harvey92, harvey01}.
Therefore, we linked the amplitude and length of the ephemeral region cycles to those of the corresponding sunspot cycles based on the R$_{g}$ record.
The relationships that provide the link contain two free parameters: the amplitude, $X$, and the time extension, $c_{x}$, of the ephemeral region cycles \citep[for more details see][]{Dasi-Espuig14}. 

The values of $X$ and $c_x$ were kept constant over the whole period considered in this study, and therefore we assume that the dynamo has behaved in the same way since 1700. Between about 1650 to 1700, the Sun was in a state of a grand minimum (Maunder minimum), when extremely few sunspots were observed \citep{eddy76, hoyt98, usoskin16}. Our model, therefore, does not produce ephemeral regions at this time as they are linked to the sunspot cycle. $^{10}$Be cosmogenic isotope data have indicated that the solar cycle continued throughout the Maunder minimum, albeit in too weak a manner to produce sunspots, and thus producing only faculae and ephemeral regions \citep{beer98, usoskin16}. If true, the relationships used here may not be valid during times of grand minima, and would mean that the reconstructed irradiance is too low between about 1700 and 1715, i.e. the change in the TSI since the Maunder Minimum obtained with this model would be an upper limit. On the other hand, we have assumed that for a given magnetic flux, network elements (including ephemeral active regions) have the same brightness contrast to the quiet Sun as active region faculae. If the network elements turn out to be brighter \citep[e.g.][]{solanki86, ortiz02, yeo13}, then the change in TSI since the Maunder minimum would be larger than the value obtained here.

Note that the monthly smoothed sunspot number during cycle 24 reached a maximum of 81.9 in April 2014\footnote{http://solarscience.msfc.nasa.gov/predict.shtml}. We have updated the present study \citep[with respect to][]{Dasi-Espuig14} to obtain the magnetic flux from ephemeral regions corresponding to cycle 24, that are present during the end of cycle 23.

  \section{Irradiance}

The SATIRE-T2 model classifies the photospheric magnetic features into 4 main classes: sunspot umbra ($u$), sunspot penumbra ($p$), active region faculae ($f$), and network produced in the ephemeral regions ($e$). Regions free of magnetic fields correspond to the quiet sun ($q$).
The irradiance of each component is given by its corresponding intensity spectrum and the fraction of the solar surface it covers.

The intensities are time-independent, but depend on the wavelength and the heliocentric angle $\theta$ (where $\mu=\cos\theta$). 
We employed the intensities $I(\mu,\lambda)$ computed by \cite{unruh99}, who used the spectral synthesis code ATLAS9 of \cite{kurucz92}. The sunspot umbrae and penumbrae are described by radiative equilibrium models \citep{kurucz91} with effective temperatures of 4500 K and 5450 K, respectively; the quiet sun is represented by the standard solar model atmosphere of \cite{kurucz92},
 and the faculae by a modified version of FAL-P \citep{fontenla93}.

The fraction of the solar surface covered by each component varies with time as the photospheric magnetic field evolves. 
The area coverage of sunspots and faculae is obtained from the semi-synthetic records (see Sect. 3.1) and the simulated magnetograms, respectively, and is given by spatially-resolved filling factors. These filling factors refer to the fraction of all pixels within limb intervals of $[\mu - \Delta \mu/2, \mu + \Delta \mu/2]$ ($\Delta \mu$ is typically 0.01) that are filled by a particular magnetic feature. They are denoted by $\alpha(\mu, t)$.
For the ephemeral regions we used 
the (disc-integrated) area coverage, 
$f^{e}(t)$, since their magnetic flux in our model was provided separately from the SFTM, for the whole disc (see Sect. 2.2).

We calculated the spectral irradiance as follows:
\begin{eqnarray}
  S(t,\lambda)&=&\sum_{\mu} \bigg[ 
    \alpha^{u}(\mu,t) I^{u}(\mu,\lambda) 
 + \alpha^{p}(\mu,t) I^{p}(\mu,\lambda) \nonumber \\
&& + \alpha^{f}(\mu,t) I^{f}(\mu,\lambda)
 + \alpha^{q}(\mu,t) I^{q}(\mu,\lambda)
    \bigg] \\
 &&+ f^{e}(t) F^{f}(\lambda) - f^{e}(t)F^{q}(\lambda) \nonumber,
\end{eqnarray}
where $F^{f}(\lambda)$ and $F^{q}(\lambda)$ correspond to the disc-integrated intensities of faculae and the quiet sun, respectively (see Sect. 3.1), and $\sum_{\mu}$ indicates the sum over all pixels on the solar disc. 

\begin{figure}
  \includegraphics[width=0.5\textwidth]{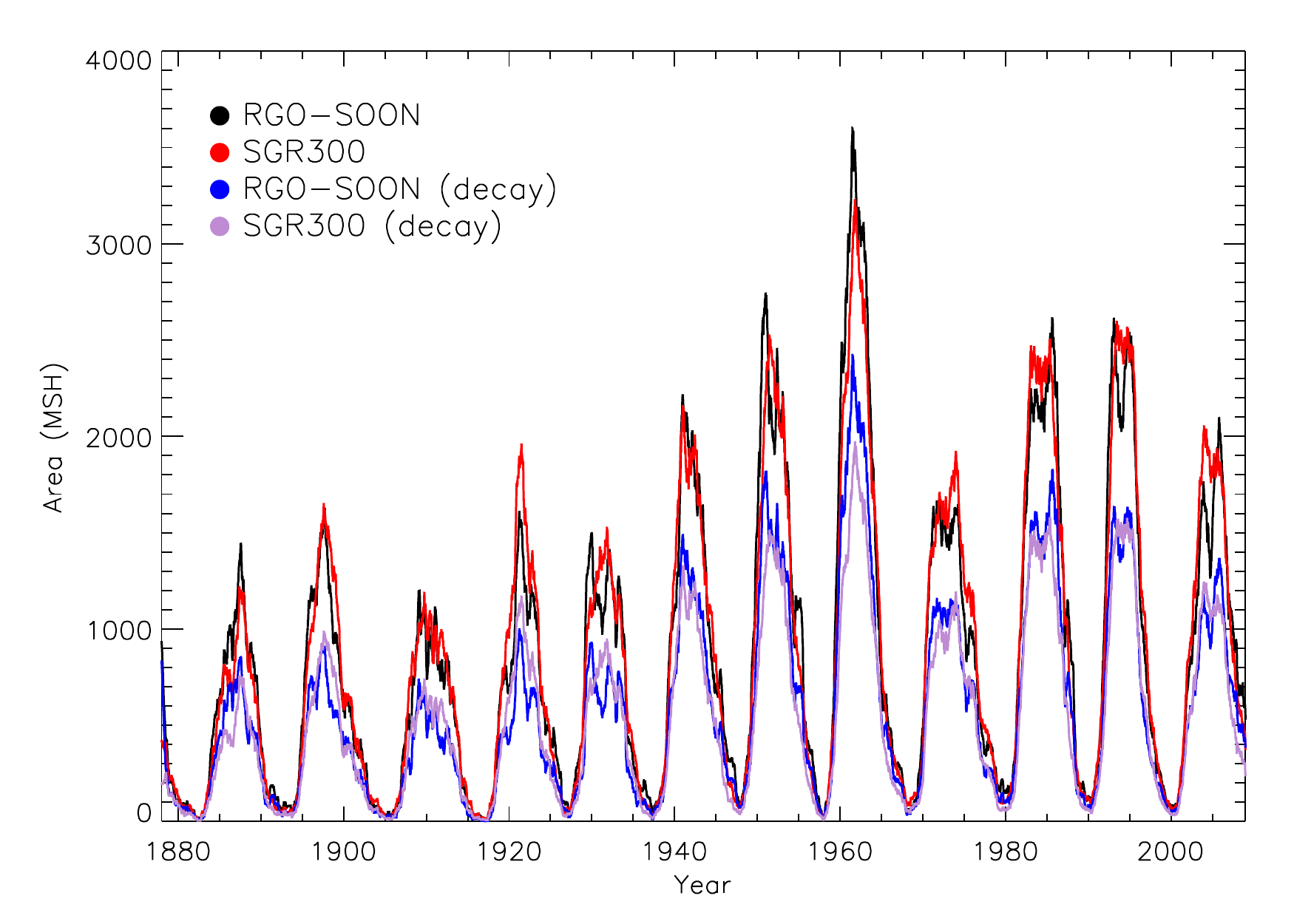}
  \caption{Daily sunspot group area on the visible side of the sun as recorded in the RGO-SOON data set (black), and that obtained from the synthetic record SGR300 within $\pm70^{\circ}$ from the central meridian (red). The same is plotted but taking only the decay phase of sunspot groups in the RGO-SOON data set (blue) and SGR300 (magenta).}
 \label{ss_area}
\end{figure}

 \subsection{The filling factors}

The semi-synthetic records contain the position, area, and tilt angle of sunspot groups only on the day of maximum area. This information is enough to model the photospheric magnetic flux with the SFTM \citep[e.g.]{cameron10, jiang11}. However, to reconstruct irradiance we also need the area and position of each group on all days on which they are visible on the solar disc.
To compute sunspot filling factors, i.e. the darkening caused by sunspots, we therefore extend the semi-synthetic records to include the days before and after a group reaches its maximum area by using empirical relationships, as described below. Note that we assume that the brightness of sunspot umbrae and penumbrae does not change in the course of their evolution.

To obtain the latitude and longitude of the sunspot groups on the days before and after the time of maximum area we used the observed profiles of meridional circulation from \cite{vanBallegooijen98} and differential rotation from \cite{snodgrass83}, consistent with the SFTM.

Since the exact form of the decay law of sunspot groups is uncertain, we tested different decay laws and rates. In the literature we found studies that support both linear and parabolic decay laws, as well as different values of the decay rates \citep{morenoinsertis88, martinezpillet97, petrovay97, baumann05, hathaway08}. We selected the studies that dealt with the decay of total (umbral plus penumbral) sunspot group area instead of individual spots as the semi-synthetic records list total sunspot group area.

\citet{hathaway08} used the combined RGO and Solar Optical Observing Network (SOON) data sets covering the periods 1974 to 1976, and 1977 to the present, respectively, to obtain an instantaneous decay rate as a function of group area, $A_{s}$,
\begin{equation}
 \varGamma(A_{s}) = (24 \; \rm{ MSH} + 0.116 A_{s}) \; \rm{/day}, 
\end{equation}
where the area is measured in millionths of the solar hemisphere (MSH). The function $\varGamma$ is defined as the average of the difference in area between three consecutive days: $ \varGamma(A_{s,i}) = [A_{s,i-1} - A_{s,i+1}]/2$, where $i$ denotes the day number.
In contrast, \cite{baumann05} modeled the distribution of sunspot group areas employing a linear and a parabolic decay law for different constant decay rates.
Given that the difference between the areas predicted by the instantaneous decay rate as a function of area of \cite{hathaway08} and the parabolic decay law of \cite{baumann05} is minimal, we computed areas with Eq. (2). The correlation coefficient between the observed and predicted daily sunspot areas smoothed over a year for the decay phase of sunspot groups is r = 0.98. 

For the growth phase of sunspot groups we used the RGO-SOON record\footnote{http://solarscience.msfc.nasa.gov/greenwch.shtml} to derive instantaneous growth rates as a function of maximum area. We calculated the growth rate as $g = (A_{s,max}-A_{s,0})/(t_{max}-t_{0})$, where $A_{s,max}$ and $t_{max}$ denote the maximum area and the time of maximum area respectively, and $A_{s,0}$ and $t_{0}$ denote the area and date at the time of first observation respectively. A linear fit of the growth rate versus $A_{s,max}$ gave a slope of 0.163/day. This value was used to obtain the corresponding growth rate for each sunspot group in the semi-synthetic record (according to its given maximum area).
The correlation coefficient between the observed and predicted daily sunspot areas smoothed over a year for the growth phase of sunspot groups is r = 0.96. 

We extended the semi-synthetic sunspot group record (SGR) by including the growth and decay phases of each sunspot group before and after the day each group reaches maximum area, respectively. 

Since the semi-synthetic sunspot records are of statistical nature, we constructed two different ones (hereafter referred to as SGR300 and SGR301) that fulfil the statistical relationships found by \cite{jiang11-a} in order to test the sensitivity of our reconstructions to the choice of the semi-synthetic record.
The differences between the two datasets result from the different random numbers in the generation of the sunspot group emergences based on the relationship between the statistical properties of sunspot emergence and the cycle phase and strength \citep{jiang11-a}.

Figure~\ref{ss_area} shows the daily sunspot area smoothed over a year from the SGR300 (red) and the RGO-SOON (black) records.
The correlation coefficient between the observed and predicted daily sunspot areas smoothed over a year for the full data set is r = 0.98. Although not shown in Fig.~\ref{ss_area}, the daily sunspot group area obtained from SGR301 is very similar to SGR300, with a correlation coefficient between the observed and predicted yearly smoothed areas of r = 0.98.
To demonstrate the good match with the decay and growth phases (i.e. not only with the full data set), we added in Fig.~\ref{ss_area} the daily sunspot area corresponding to the decay phase of groups from observations (blue) and SGR300 (magenta).  

We retrieved the location and area of the sunspot groups from the extended semi-synthetic records and masked the corresponding pixels in the simulated magnetograms. The remaining pixels in the simulated magnetograms were used to compute the filling factors of faculae and the quiet Sun following the method described in \cite{Dasi-Espuig14}. 

We identified the pixels filled by sunspots and computed the corresponding filling factors, $\alpha^{s}(\mu, t)$, for every $\mu$ bin.
The corresponding filling factors of sunspot umbra, $\alpha^{u}(\mu, t)$, and penumbra, $\alpha^{p}(\mu, t)$, were obtained assuming the empirical ratio of umbral to total area, $A_{u}/(A_{u}+A_{p}) = 0.2$ \citep{brandt90, wenzler05}.

The pixels that are not covered with sunspots were considered to be covered by faculae and quiet sun atmospheres. 
To obtain the facular filling factor we first computed histograms of the total fraction of such pixels within $[\mu - \Delta\mu/2, \mu + \Delta\mu/2]$ and $[B - \Delta B/2, B + \Delta B/2]$ ($\Delta B \sim 2 G)$, and then multiplied each $B$ bin by the fraction $B/B_{sat}^f$ until the saturation value $B^{f}_{sat}$ is reached. Summation over all $B$ bins produces the facular filling factor for every $\mu$, $\alpha^{f}(\mu, t)$.
For $B$ bins below the saturation value we still have a portion of the $\mu$ bin that corresponds to the quiet sun and is equal to $\alpha^{q}(\mu, t) = N(\mu)-\alpha^{u}(\mu, t)-\alpha^{p}(\mu, t)-\alpha^{f}(\mu, t)$, where $N(\mu)$ is the number of pixels in each $\mu$ bin.

For ephemeral regions we employed the disc-integrated magnetic fluxes. The filling factors of ephemeral regions, $f^{e}(t)$, are also considered to be proportional to their corresponding magnetic flux density ($B^{er}(t)$) as $B^{er}(t)/B^{er}_{sat}$.

\subsection{Parameters and model validation}
\label{S:params}
 \begin{table}
 \begin{minipage}{\columnwidth}
 \caption{Free parameters of the model that provide the best fit to the measured total photospheric magnetic flux and the total solar irradiance. The last column lists the uncertainty range of each free parameter (see Sect. 3.2).}
 \centering
 \small
 \begin{tabular}{l c c c}
 \hline\
   Parameter  &  Notation  &  Value  &  Ranges \\
 \hline
 AR scaling factor			&  $a$ & 0.74 & 0.70 -- 0.85 \\ 
 ER amplitude factor 		&  $X$ & 0.69 & 0.45 -- 0.75 \\
 ER cycle extension [years] 	&  $c_{x}$ & 7.48 & 7.45 -- 7.50 \\
 Saturation flux [G] 			&  $B_{sat}$ & 456 & 450 -- 500 \\
 \hline
 \end{tabular}
 \end{minipage}
\end{table}

The SATIRE-T2 model presented in \cite{Dasi-Espuig14} has 5 free parameters.
These are the scaling factor of the total simulated magnetic flux in active regions, $a$; the amplitude, $X$, and time extension, $c_{x}$, of the ephemeral region cycle with respect to the active region cycle; the saturation value for the faculae $B^{f}_{sat}$ and ephemeral regions, $B^{er}_{sat}$. The factor $a\in(0,1)$ was introduced in \cite{Dasi-Espuig14} to re-scale the free parameter $B_{max}$ in the SFTM. $B_{max}$ is the peak field strength of each active region at the time of emergence and is fixed in the SFTM such that the absolute values of the simulated total magnetic flux match the measurements from the Mount Wilson (MWO) and Wilcox Solar (WSO) observatories \citep[see also][]{cameron10, jiang11}. The measured total magnetic flux consists of magnetic flux emerging within active and ephemeral regions, but the SFTM does not include the latter mainly due to the lack of direct and continuous observations of ephemeral regions.
Therefore, the scaling factor $a$ is needed when combining the magnetic flux from the SFTM and that of the ephemeral regions as described in Sect. 2.2.

As in \cite{Dasi-Espuig14}, we fixed the values of the free parameters using the optimisation routine PIKAIA\footnote{http://www.hao.ucar.edu/modeling/pikaia/pikaia.php} \citep{charbonneau95} to minimise the reduced chi-square between the observed and modeled TSI and total magnetic flux ($\chi_r^2 = \chi^2_{r,TF}+\chi^2_{r,TSI}$), over cycles 21 -- 23.
The modeled data are based on the RGO-SOON sunspot group data set.
The main difference to our previous approach is that we now include the growth phase of sunspot groups to compute daily filling factors, rather than just their decay phase. This improves the irradiance reconstructions on daily and monthly timescales and provides filling factors that are more consistent with the SATIRE-S model based on measured magnetograms \citep{yeo14}, which is key to obtaining the correct spectral response (see Sect. 4.2).

The computed TSI was fitted to the PMOD (Physikalisch-Meteorologisches Observatorium Davos)\footnote{Version d41\_62\_1302} composite of observations.
For the magnetic flux, since we are using two different records (MWO and WSO), we fitted to the average of the two.
We took into account that the measurements of the total magnetic flux miss a significant part of the magnetic flux in ephemeral regions \citep[due to the limited spatial resolution of the magnetograms;][]{krivova04} by reducing the modeled ephemeral region magnetic flux by a factor of 0.4 \citep{krivova07, krivova10}.
The total magnetic flux that we use to compare to the observations is $a\Phi^{AR}$+0.4$\Phi^{ER}$, where $\Phi^{AR}$ and $\Phi^{ER}$ represent the modeled magnetic flux from active and ephemeral regions respectively.

Here we employed 27-day averages to compute the reduced chi-square values ($\chi^2_r$) for the TSI and the total magnetic flux. None of the data sets provide errors for individual data points and thus we took the standard deviation of the measured time series as an estimate of the errors \citep[see also][]{krivova07, krivova10, vieira10}.

The resulting values of $a$, $X$, $c_{x}$, and $B_{sat}$ that optimise the TSI and the average total magnetic flux are listed in Table 1. Note that we adopted a single saturation value, $B_{sat}$, as the best-fit values for $B^{f}_{sat}$ and $B^{e}_{sat}$ agree to within 20 G.

\begin{figure}
  \includegraphics[width=0.5\textwidth]{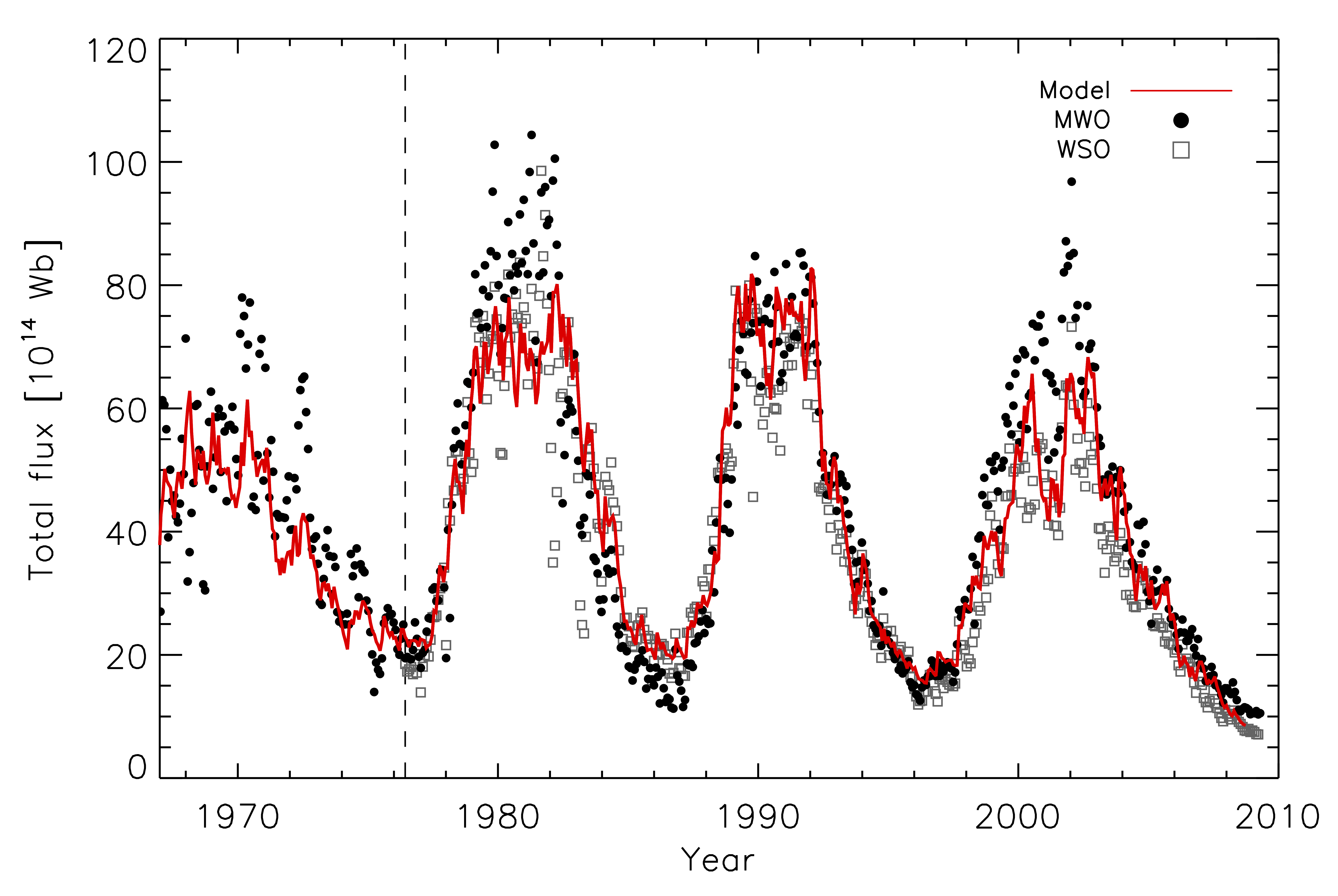}
    \caption{Comparison of the total magnetic flux measured by the Mount Wilson (filled circles) and Wilcox Solar (open squares) observatories, and modeled (red) as $a\Phi^{AR} + 0.4\Phi^{ER}$ (with a = 0.74; see Table 1) using SATIRE-T2 with the RGO-SOON sunspot group record.}
   \label{TF_obs}
\end{figure}
\begin{figure}[t]
  \includegraphics[width=0.5\textwidth]{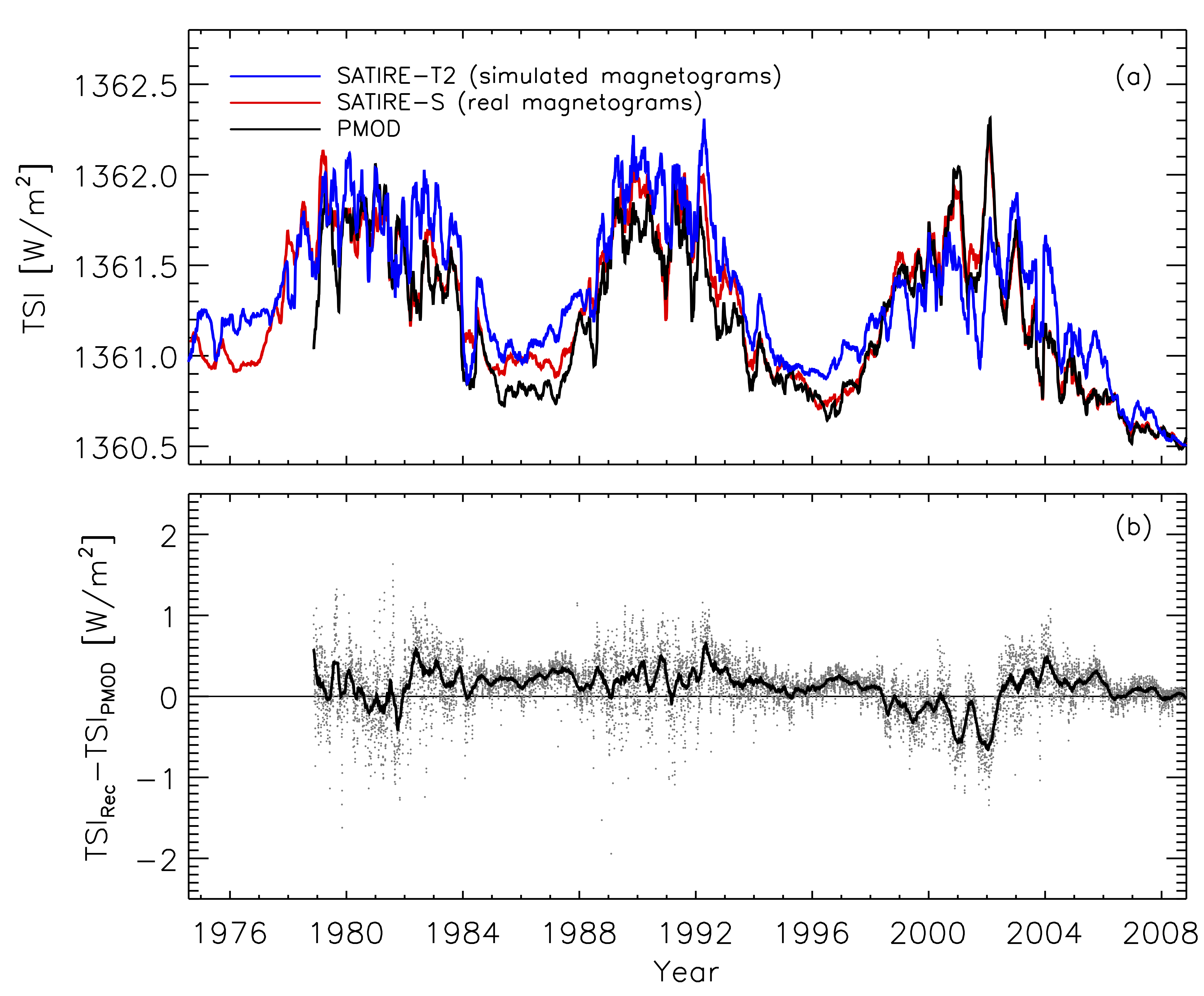}
    \caption{a) TSI reconstructed with SATIRE-T2 based on RGO-SOON data (blue), with SATIRE-S (red), and observed (PMOD composite, black). b) The difference between the TSI observed and reconstructed with SATIRE-T2. Each dot corresponds to a daily value, and the thick curves are three month running means.}
 \label{TSI_obs}
\end{figure}
\begin{figure}
  \includegraphics[width=0.5\textwidth]{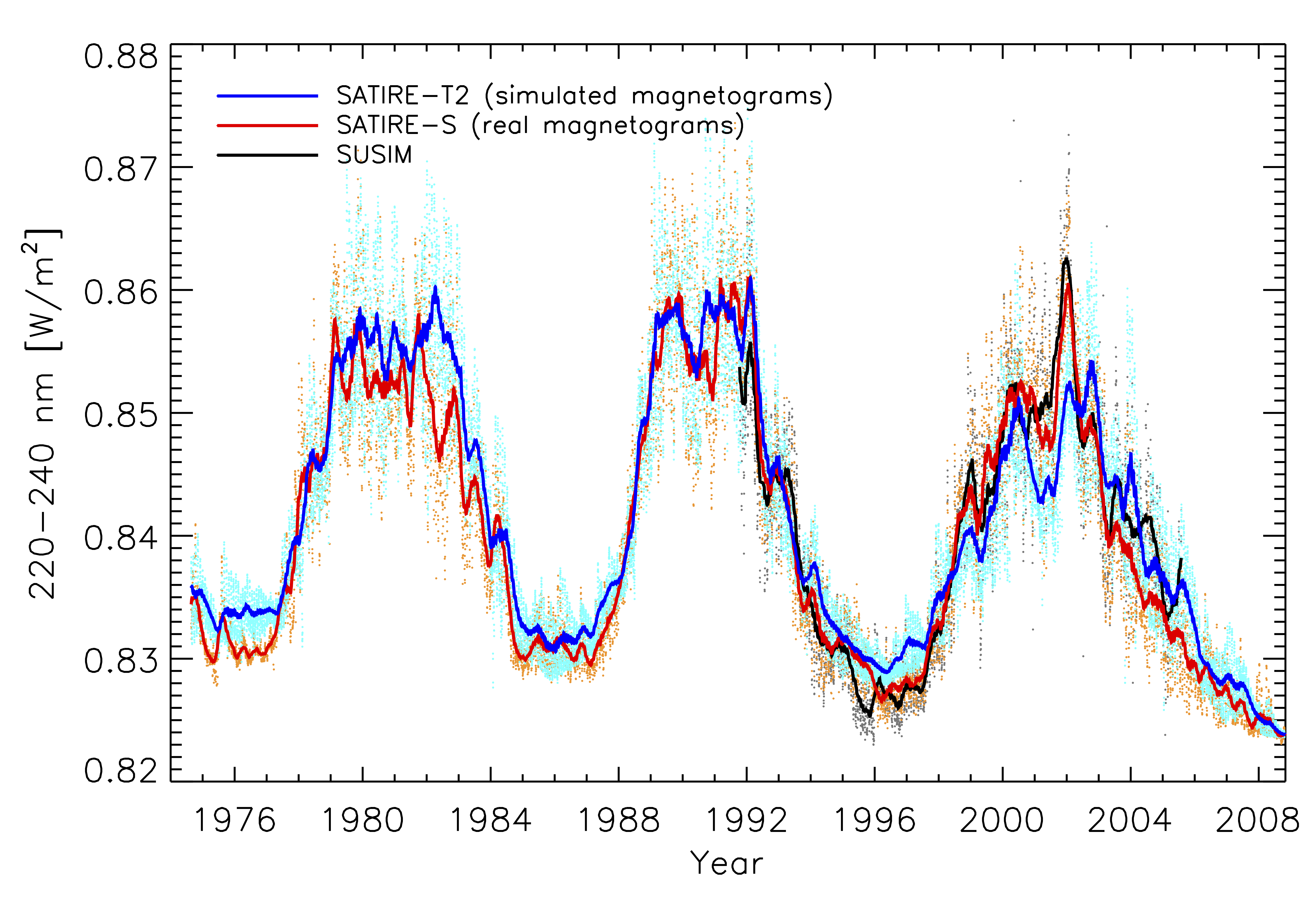}
    \caption{ Spectral irradiance between 220 nm and 240 nm reconstructed with SATIRE-T2 based on RGO-SOON data (blue), with SATIRE-S (red), and measured (SUSIM, black). Each dot corresponds to a daily value, and the thick curves are three month running means.}
 \label{UV_obs}
\end{figure}

The best-fit and observed total magnetic flux are presented in Fig.~\ref{TF_obs}. The modeled total magnetic flux based on RGO-SOON sunspot group data is represented by the thick red line, whereas the measurements taken by the Mount Wilson (MWO) and Wilcox Solar (WSO) observatories \citep{arge02, wang02-b, wang06} are indicated, respectively, by the filled circles and open squares.
The measurements are provided for each Carrington rotation, which corresponds to a period of approximately 27 days. 
The correlation coefficient (r), slope ($s$), and reduced chi-square ($\chi_{r, TF}^{2}$) of a linear fit between the average of the measurements from the two observatories, and the modeled total magnetic flux are, respectively, r = 0.95, s = 1.01, and $\chi_{r,TF}^{2}$ = 0.093 for the 27-day averages (see Table~\ref{T:fits}). For the fit we only considered the period when the measurements from both observatories overlap, i.e. after 1976. We nevertheless plotted the data from the MWO from 1967 on, i.e. from when they became available, to show the good correspondence of the model to these.

The comparison between the modeled and observed TSI is displayed in Fig.~\ref{TSI_obs} between November 1978 and the end of 2008. Figure~\ref{TSI_obs}a shows the TSI reconstructed with SATIRE-T2 based on the RGO-SOON record in blue, with SATIRE-S in red, and the PMOD composite in black. The rms deviation, $\Delta_{rms}$, between the monthly averages of our reconstruction and that of the SATIRE-S is smaller ($\Delta_{rms}$ = 0.31 Wm$^{-2}$) than that between the SATIRE-S and the ACRIM \citep[$\Delta_{rms}$ = 0.40 Wm$^{-2}$;][]{willson03} and IRMB \citep[$\Delta_{rms}$ = 0.41 Wm$^{-2}$;][]{dewitte04} composites \citep[][not plotted]{yeo14}. 
Both the PMOD composite and the SATIRE-T2 reconstruction have been shifted to the average TSI in 2008 measured by SORCE/TIM \citep{kopp05-a, kopp05-b, kopp05-c}.
In Fig.~\ref{TSI_obs}b we plot the difference between the SATIRE-T2 reconstruction and the observations. According to \cite{frohlich09}, the typical error of the PMOD composite during the three minima can be up to 0.3Wm$^{-2}$. 

The fit parameters between our reconstruction and the PMOD composite, i.e. the slope, correlation coefficient, and reduced chi-square, for the 27-day averages are $s$ = 0.87, r = 0.87, and $\chi_{r, TSI}^{2}$ = 0.257. In Table 2  we also listed the fit parameters for daily and 3-month timescales.
The penultimate column provides the reduced chi-square weighted by a function that reduces the influence of outliers \citep[for details see][]{vieira10}, in order to compare our fits to those obtained by \cite{krivova10} with the SATIRE-T model (see last column).

As discussed in \cite{Dasi-Espuig14}, the SATIRE-T2 model reproduces the total magnetic flux and the TSI on monthly timescales more accurately than SATIRE-T due to (1) a more realistic description of the evolution of the facular field and (2) the use of spatially resolved magnetograms. 
We also note that, thanks to the incorporation of the growth phase of sunspot groups, the correlation coefficient and the reduced chi-square value of the TSI on daily timescales have improved significantly compared to those quoted by \cite{Dasi-Espuig14}, who had r = 0.78 and $\chi_{r, TSI}^{2}$ = 0.413.

Figure~\ref{UV_obs} shows the spectral irradiance integrated over the wavelength range 220 -- 240 nm, reconstructed using the SATIRE-T2 (blue) and SATIRE-S (red) models, as well as measured by UARS/SUSIM \citep{brueckner93, floyd03-b}. 
For comparison, the UARS/SUSIM observations have been offset to match the SATIRE-S minimum levels in the year 1996.
The good match between our reconstruction and the UARS/SUSIM measurements, as well as with the SATIRE-S model, at this wavelength range is encouraging, given the fact that our model is optimised to fit only the observations of TSI and total magnetic flux.

The low values of $\chi_{r, TF}^{2}$ = 0.093 and $\chi_{r, TSI}^{2}$ = 0.257 for the total magnetic flux and the TSI respectively, suggest that we overestimated the errors associated with the measurements. 
Rescaling the errors for the total magnetic flux so as to force the reduce chi-square value to be of the order of 1 implies $\sigma \approx$ 2$\times10^{14}$Wb. This is significantly below the typical discrepancy between the MWO and WSO measurements of 5 -- 10$\times10^{14}$Wb, especially during the 4 years centred around solar maximum. We thus refrained from rescaling, but compared the reconstructions that result when fitting each record separately (see Sect. 3.3).

In principle it is possible to use constant chi-square boundaries to estimate confidence limits on the parameters \citep{press92}.
For 4 free parameters $\Delta\chi^2 = 16.3$ corresponds to a 3-$\sigma$ confidence limit, where $\Delta\chi^2 = \chi^2 - \chi_{min}^2$ and $\chi_{min}^2$ is the minimum chi-square.
We warn, however, that due to our crude error estimates and our low reduced chi-square values, the true confidence interval is likely to be lower (a rough scaling suggests it to be 
more representative of 1-$\sigma$).

To gauge the effect of uncertainties in the parameter fitting, we calculated $\Delta\chi^2$ on a grid of parameters within the allowed ranges: we vary $a$ between 0.5 -- 1.0 in steps of 0.05, $X$ between 0.45 -- 1.2 in steps of 0.05, $c_x$ between 5 -- 7.5 years in steps of 0.5 years, and B$_{sat}$ between 200 -- 700 G in steps of 50 G \citep[for more detail on the allowed ranges see][]{Dasi-Espuig14}. All sets of parameters that fall within $\Delta \chi^2 \leq 16.3$ (see Table 1 for the ranges)
are used to reconstruct the TSI back in time and used to provide an estimate of the uncertainty resulting from the parameter fitting (see Sect.~3.3 for further discussion).

\begin{table}[t]
 \begin{minipage}{\columnwidth}
 \centering
 \small
 \renewcommand{\footnoterule}{}  
 \caption{Slope, correlation coefficient (r), and reduced chi-square ($\chi_{r}^{2}$) between the modeled and observed total magnetic flux (TMF) and TSI, and between the modeled UV (220 -- 240 nm) irradiance with SATIRE-S and SATIRE-T2, for different averaging periods. Note that only the total magnetic flux and the TSI are used to optimise the SATIRE-T2 model.}
 \begin{tabular}{l l c c c c c }
 \hline\
   Quantity  &  Scale  &  Slope  &  r  &  $\chi_{r}^{2}$  &  $\chi_{r,w}^{2}$ \footnote{Reduced chi-square of SATIRE-T2 following the definition used by \cite{krivova10} } 
   											    &   $\chi_{r,w}^{2}$ \footnote{Reduced chi-square of the SATIRE-T model } \\
 \hline
 TSI			        &  1 day        &  0.82 & 0.86  &  0.331  &  0.218 &  0.233 \\ 
 UV		 		&  1 day        &  0.96 & 0.92  &  0.159  &  0.098 &  - \\
 \hline
 TMF		 	&  27 days \footnote{Timescale employed to optimise the model}
                                  		    &  1.01 & 0.95  &  0.093  &  0.032  &  0.069 \\
 TSI			 	&  27 days    &  0.87 & 0.87  &  0.257  &  0.156  &  - \\
 UV		 		&  27 days    &  1.00 & 0.95  &  0.089  &  0.053  &  - \\
 \hline
 TSI			 	&  3 months  &  0.89 & 0.88  &  0.320  &  0.214  &  - \\
 UV		 		&  3 months  &  1.00 & 0.96  &  0.075  &  0.045  &  0.072 \\
 \hline
 \end{tabular}
 \end{minipage}
\label{T:fits}
\end{table}
 \begin{figure*}
  \includegraphics[width=0.5\textwidth]{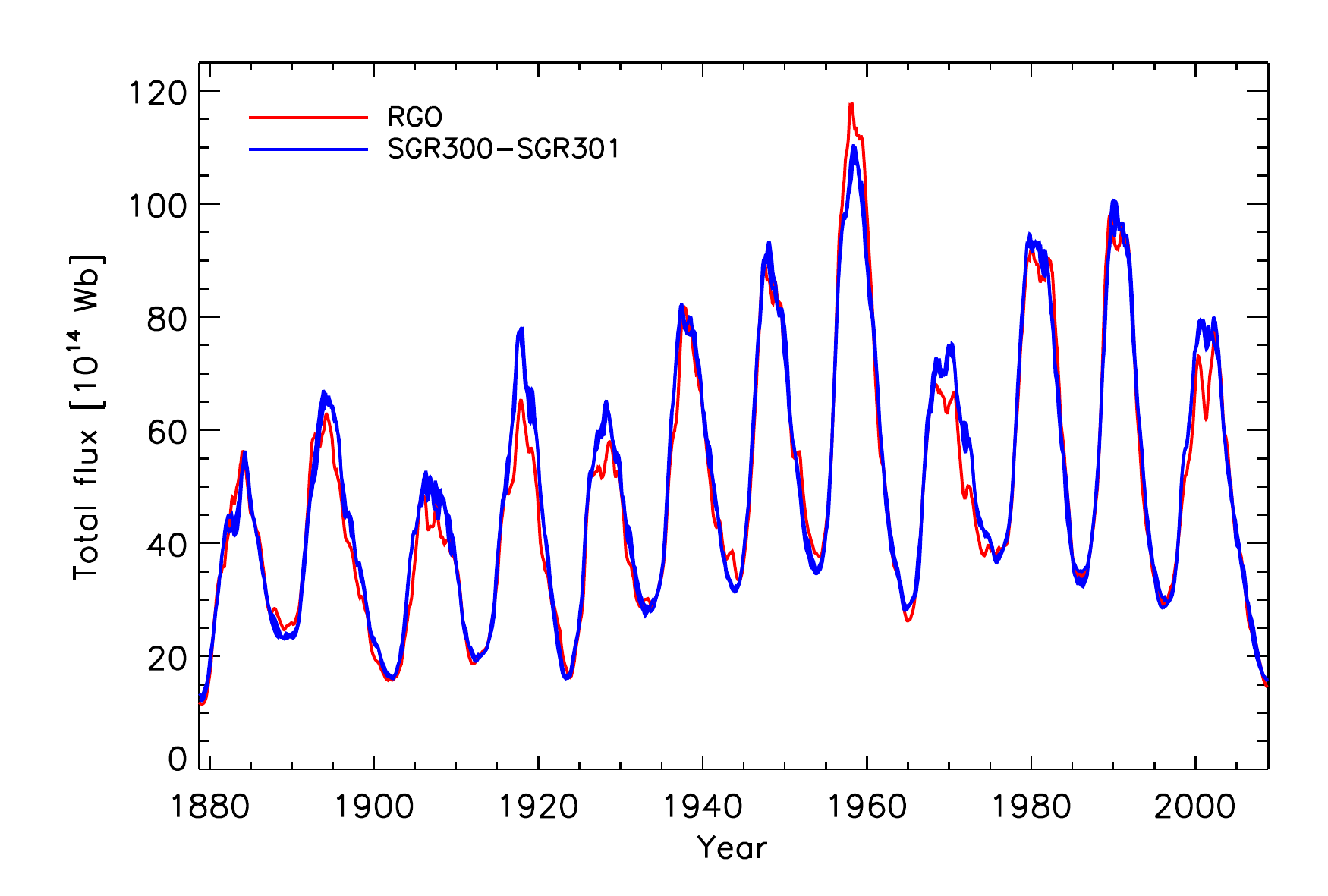}
  \includegraphics[width=0.5\textwidth]{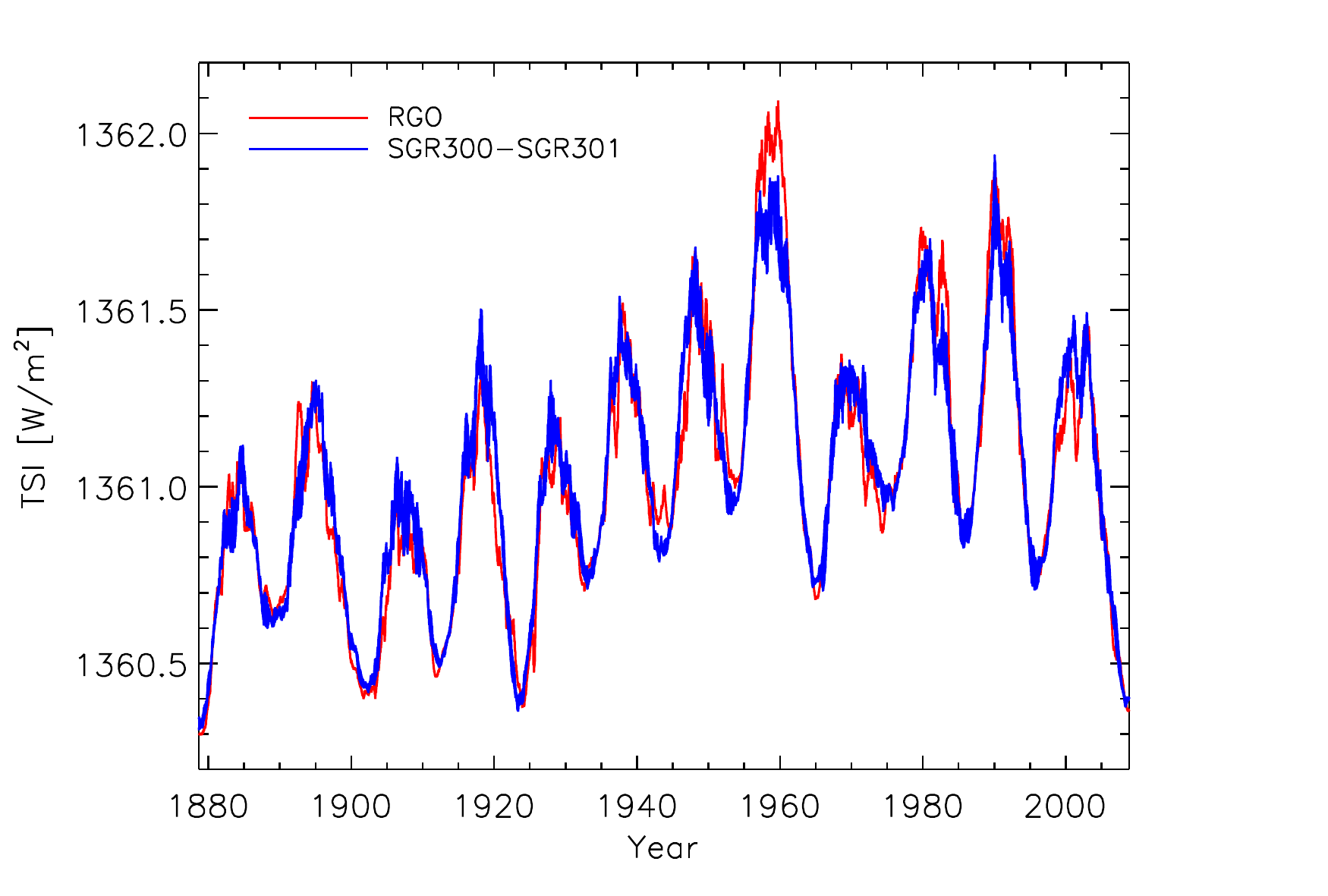}
    \caption{Comparison between the 1-year smoothed total magnetic flux (left) and TSI (right) reconstructions based on the RGO-SOON sunspot group record (red), and the semi-synthetic sunspot group record (blue). The thickness of the blue line indicates the range of values between two reconstructions based on the semi-synthetic records SGR300 and SGR301.}
 \label{SSI_RGO_synth}
\end{figure*}

 \subsection{SATIRE-T2 extension back to 1700}

In Sect. 2.1 we extended the RGO-SOON sunspot group record back to 1700 employing the semi-synthetic data sets of \cite{jiang11-a}. We computed daily simulated magnetograms using the semi-synthetic data sets as input into the SFTM, and daily sunspot filling factors using SGR300 and SGR301 (as described Sect. 3.1), to reconstruct the spectral solar irradiance over the past 300 years. For this we used the values of the free parameters listed in Table 1. Note that the choice of the initial field strength in the SFTM may affect the first $\sim$20 years of the modeled active region flux. In our case we set it to zero since the average value of R$_{g}$ between 1700 and 1710 is below 5 \citep{jiang11}.

This extension of the SATIRE-T2 model is suitable for irradiance reconstructions on timescales of a few solar rotations or longer since the semi-synthetic sunspot group records are built to keep the emergence rate of sunspot groups in a month proportional to the monthly sunspot group number R$_{g}$ (see Sect. 3.1). For this reason we focus on comparing the SATIRE-T2 irradiance based on RGO-SOON, SGR300 and SGR301 data sets on yearly timescales or longer.

Figure~\ref{SSI_RGO_synth} shows the yearly smoothed reconstructions of the total magnetic flux (left) and TSI (right) since 1878 based on the RGO-SOON record (red) and the semi-synthetic sunspot group records SGR300 and SGR301 (blue) using the same values for the free parameters. The thickness of the blue curves indicates the range of values between the two independent reconstructions based on SGR300 and SGR301.
The rms deviation between the yearly smoothed reconstructions based on RGO and SGR300 on the one hand, and RGO and SGR301 on the other hand, is, respectively, 4.1 and 4.2$\times10^{14}$ Wb for the total magnetic flux, and 0.094 and 0.097 Wm$^{-2}$ for the TSI.
This indicates that reconstructions based on synthetic sunspot records agree well with the one based on the RGO dataset without the need to re-fit the free parameters.   

We also tested the sensitivity of our reconstructions to the choice of the total magnetic flux data set.
If we use MWO or WSO separately to fit the free parameters in the model both the cycle amplitude and the secular increase since 1700 show differences. However, this difference is only slightly larger to that between the reconstructions using the two different semi-synthetic records SGR300 and SGR301. The cycle amplitude in the TSI varies up to 0.05 Wm$^{-2}$, while the increase between the end of the Maunder Minimum (1700) and the average over the last 3 cycles varies by 0.01 Wm$^{-2}$.

\begin{figure}
  \includegraphics[width=0.5\textwidth]{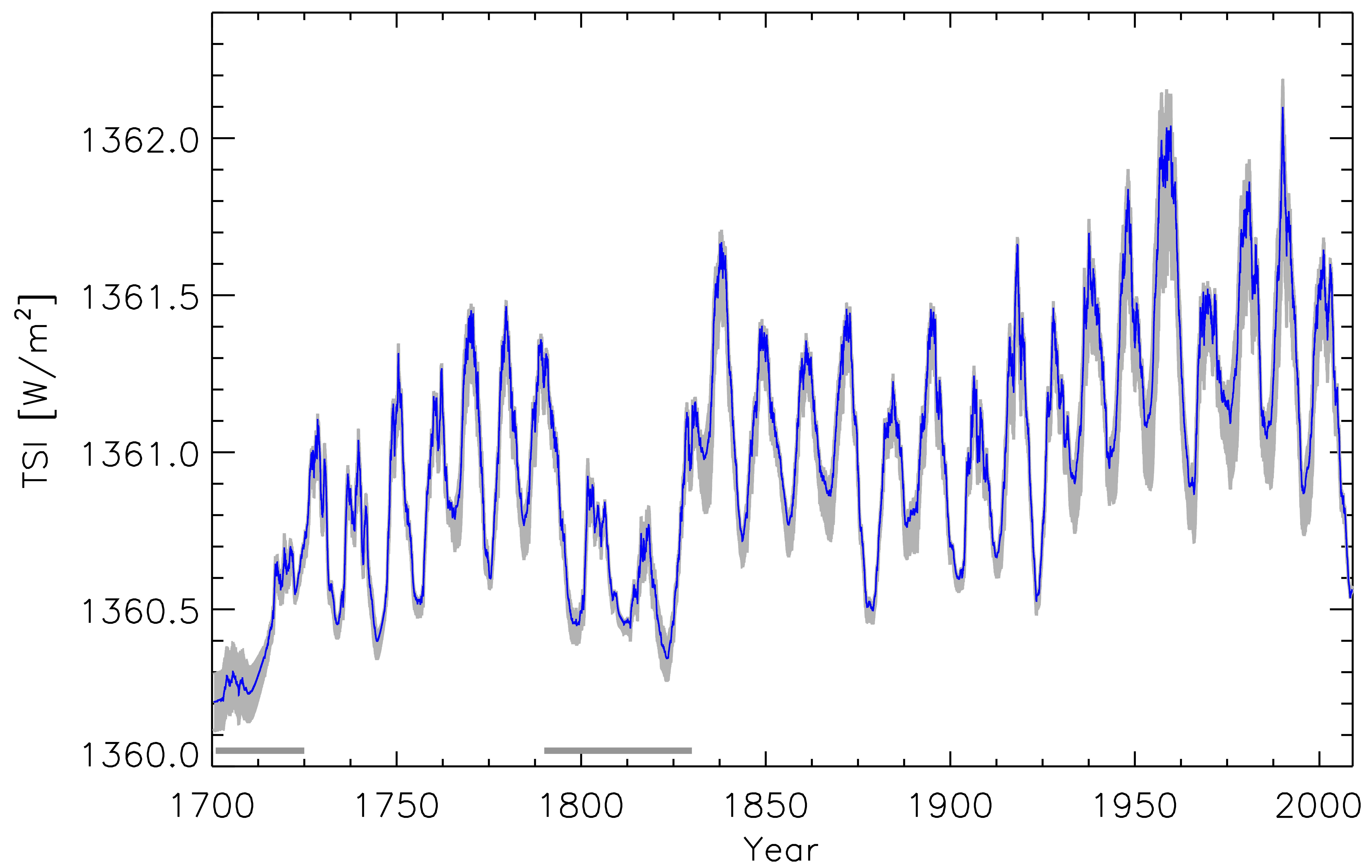}
    \caption{TSI reconstruction based on R$_g$ smoothed over 1 year (blue) and the uncertainty range (grey) based on the uncertainty in the parameter fitting (see Sect. 3.2). The grey bars at the bottom indicate the times when the statistical relationships have been extrapolated beyond the range of observations on which they are based.}
 \label{TSI_3sigma}
 \end{figure}

As discussed in Sec.~\ref{S:params}, we estimated the uncertainty associated with the parameter fitting by reconstructing the TSI back to 1700 for all sets of parameters with $\Delta\chi^2 \leq 16.3$. All of these reconstructions lie within the range shown by the grey shading in Figure~\ref{TSI_3sigma}. The blue lines shows the `best-fit' reconstruction, smoothed over one year. All of the curves have been normalised by the minimum after cycle 23 (i.e. 2007 to 2008).
The overall trend is very similar for all reconstructions. Typically, the difference in cycle amplitude remains below 0.2 Wm$^{-2}$. The uncertainty in the increase in TSI between 1700 and the average over the last three cycles
is roughly 0.5 Wm$^{-2}$.

\begin{figure}
  \includegraphics[width=0.5\textwidth]{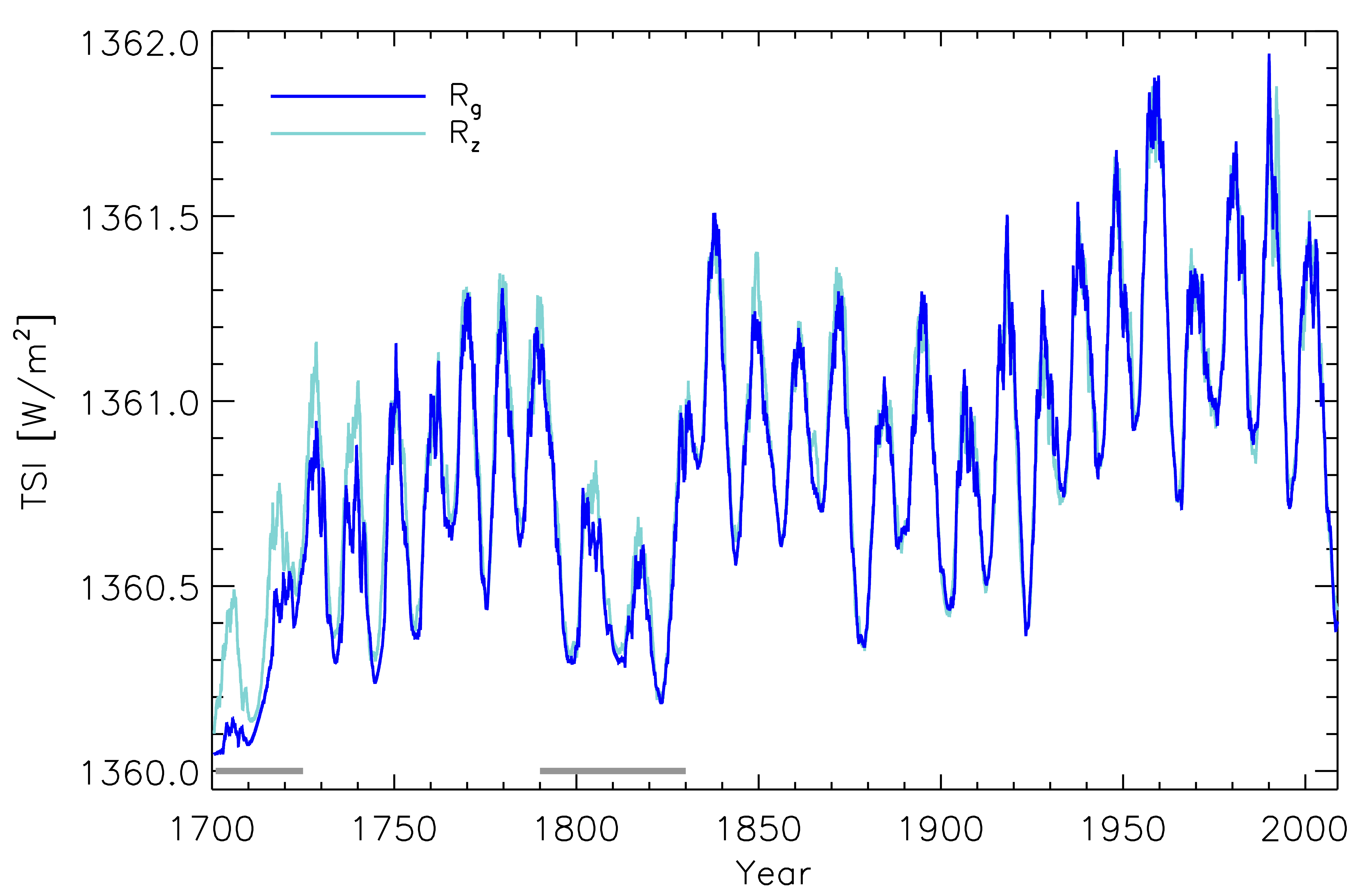}
    \caption{Comparison between the yearly-smoothed TSI reconstructions from SATIRE-T2 based on R$_g$ (dark blue) and R$_z$ (light blue). The grey bars at the bottom indicate the times when the statistical relationships have been extrapolated beyond the range of observations on which they are based.}
 \label{TSI_Rg_Rz}
 \end{figure}

Lastly, we considered differences between our reconstructions based on R$_g$ and on the Zurich sunspot number, R$_z$. Thus, we constructed a semi-synthetic sunspot group data set employing R$_z$ \citep[as described in Sect. 3.1 and in][]{jiang11-a} and repeated the steps explained in Sects. 2 and 3 to reconstruct the irradiance back to 1700 (Fig.~\ref{TSI_Rg_Rz}).
Between the years 1700 and 1800, the cycle amplitude is 0.2 -- 0.3 Wm$^{-2}$ larger for the TSI reconstruction based on R$_z$. This is because the R$_z$ is larger than the R$_g$ between 1700 and 1874. 

The calibration of the sunspot numbers is currently in a state of flux. \cite{clette14} recently published a revised R$_z$ and R$_g$ record. \cite{lockwood14} compared the R$_z$ to several other data sets to examine a possible calibration discontinuity around 1945, while additional independently corrected sunspot number series have been submitted \citep{usoskin16, svalgaard15, lockwood16}.
For this reason we decided to use in this paper the older and widely used data sets of R$_z$ and R$_g$. Once a consensus is reached as to which record is the most reliable, we will update our reconstructions.


 \section{Reconstructions back to 1700}
 
In Sect. 3.2 we validated the SATIRE-T2 model by comparing the reconstructed total magnetic flux and TSI against observations over solar cycles 21 -- 23. We also showed a good correspondence between SATIRE-T2 reconstructions of the total magnetic flux and the TSI starting in 1878 that are based on RGO-SOON data and those based on the semi-synthetic records.
In this section we discuss the total and spectral irradiance reconstructions since 1700.

 \subsection{Total solar irradiance}

\begin{figure*}
  \includegraphics[width=\textwidth]{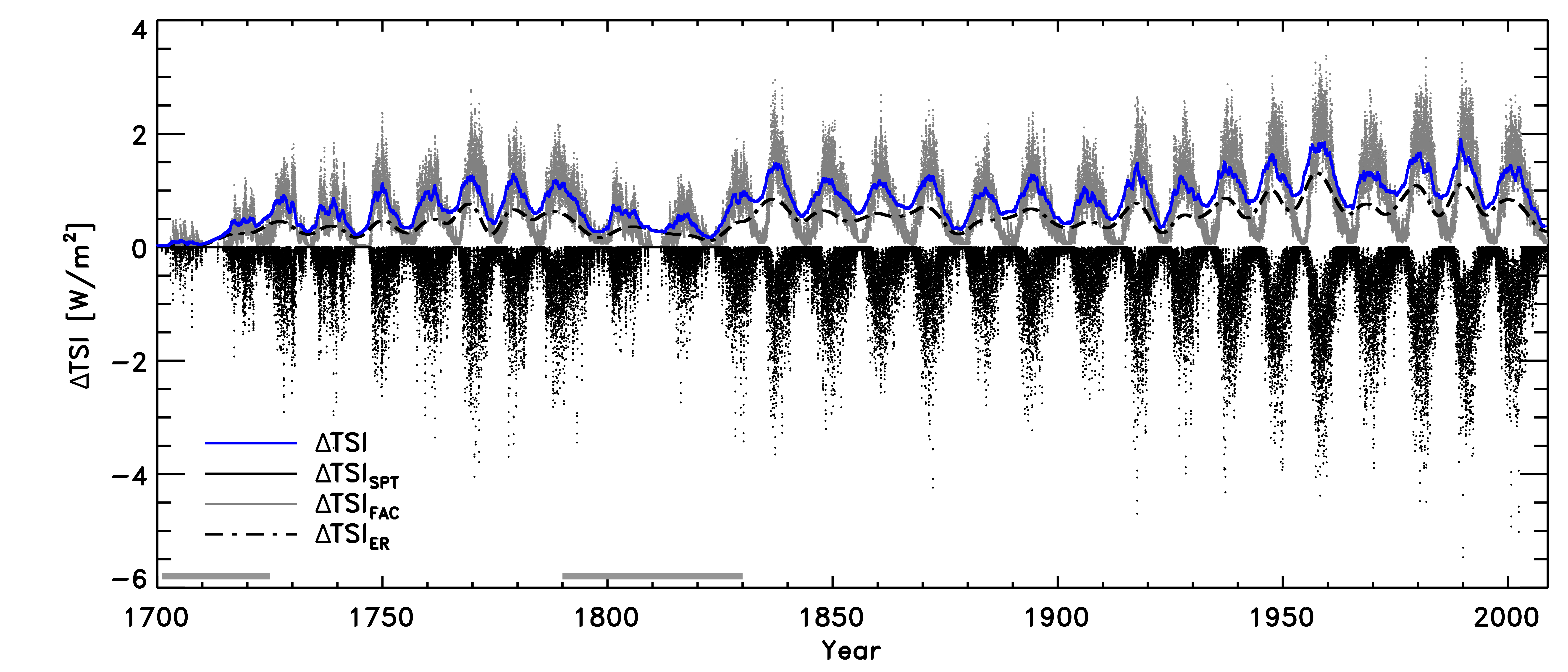}

  \includegraphics[width=\textwidth]{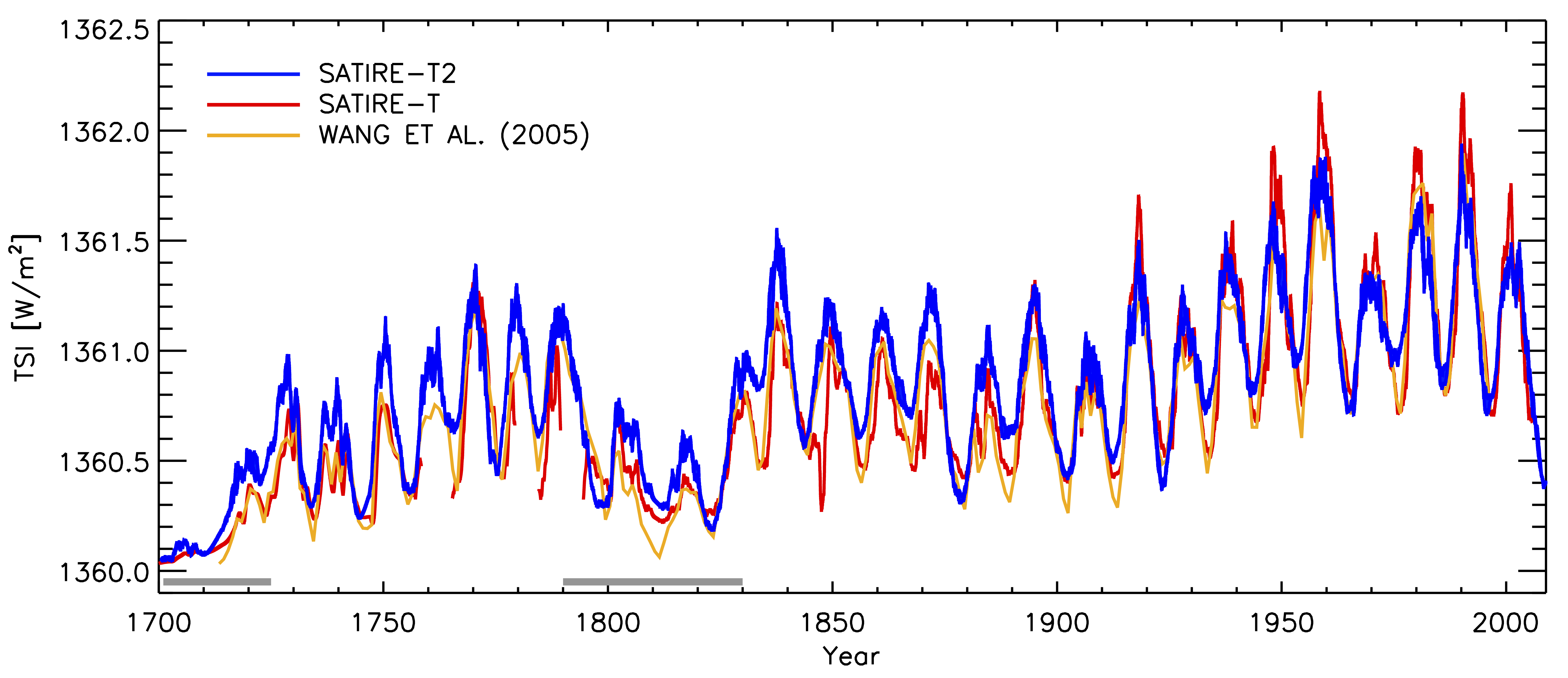}
      \caption{Top: Change in TSI relative to the quiet sun value, $\Delta$TSI, smoothed over a year (blue) based on the semi-synthetic sunspot group record SGR300. The contribution of spots, faculae, and ephemeral regions are also shown on a daily basis as grey dots, black dots and the dot-dashed black curve, respectively. Bottom: Comparison between the TSI reconstructions from SATIRE-T2 using SGR300 and SGR301 (both based on R$_g$ as input (blue), and SATIRE-T (red) smoothed over 1 year, and the yearly values of the reconstruction from \citet[orange]{wang05}. The grey bars at the bottom indicate the times when the statistical relationships of SATIRE-T2 have been extrapolated beyond the range of observations on which they are based.}
 \label{TSI_1700_comp}
\end{figure*}

The top panel of Fig.~\ref{TSI_1700_comp} shows the change in TSI relative to the quiet sun value (obtained from integrating the quiet sun model atmosphere over all wavelengths and $\mu$ values) smoothed over a year, $\Delta$TSI (blue), since 1700. The contribution of each component (spots, faculae, and ephemeral regions) is also plotted on a daily basis. 
As described by \cite{Dasi-Espuig14}, the main contributor to changes from minimum to minimum, in the SATIRE-T2 irradiance reconstructions are the ephemeral regions (see dot-dashed curve in Fig.~\ref{TSI_1700_comp}a), in agreement with previous models of the photospheric magnetic flux \citep{solanki02-b} and the SATIRE-T model \citep{krivova07, krivova10}.

We compared our TSI reconstructions with those from SATIRE-T \cite{krivova10} and \citet[hereafter WLS05]{wang05}. The bottom panel of Fig.~\ref{TSI_1700_comp} displays our yearly smoothed TSI reconstruction in blue, where the thickness of the curve represents the range of values between the two reconstructions based on the two semi-synthetic sunspot records SGR300 and SGR301. The SATIRE-T reconstruction smoothed over one year is indicated in red. Between 1750 and 1800 the reconstruction has periods with extensive gaps (due to gaps in the $R_g$) and therefore we omitted these values in the plot. WLS05's reconstruction (orange) is available only on a yearly basis, while SATIRE-T2 and SATIRE-T have a temporal resolution of one day.
To facilitate a comparison, the plotted SATIRE-T and SATIRE-T2 curves have been smoothed by a 1-year running mean.
The three reconstructions have been shifted to the average value of the PMOD composite of observations during the minimum between cycles 21 and 22, when the three reconstructions and the composite are available.

WLS05 also used a SFTM to compute the photospheric magnetic flux in active regions and their decay products from the R$_{g}$, and added the contribution of ephemeral regions separately. They, however, did not consider the extended cycles of ephemeral regions. Instead, their ephemeral region flux was taken to be constant throughout a cycle, with a value proportional to the amplitude of the corresponding sunspot cycle.
Their SFTM has also several differences to that of \cite{jiang11}. 
Firstly, they assumed a cycle-to-cycle variable meridional flow, roughly proportional to the cycle amplitude, to maintain the polarity reversals of the polar field, whereas in our case this is achieved by including the observed cycle-to-cycle variation of the sunspot group mean latitudes \citep{li03, solanki08, jiang11-a} and tilt angles \citep{Dasi-Espuig10, Dasi-Espuig13, Ivanov12, mcclintock13}. 
Note that using the statistical relationship between the tilt angles and the strength of a cycle is equivalent to taking into account the inflows toward the activity belts, i.e. variations related to the meridional flow \citep{howard80, Zhao04, gizon04, gizon05, cameron12}.
Secondly, the profile of the meridional circulation used here is in close agreement to the recent measurements of \cite{hathaway11}, while WLS05 used a profile with a sharp gradient near the equator \citep{wang89, wang09}.
Thirdly, WLS05 constructed their source function based on two synthetic records different to ours: 
in case S1 they scaled the number of emerging active regions to the yearly R$_{g}$ and fixed the strength of all regions to 5$\times10^{22}$ Mx, while in case S2 they fixed the number of emerging active regions in all cycles to 600 and scaled the strength of the active regions within a cycle to the maximum R$_{g}$ of the same cycle.
In both cases the active regions were distributed randomly in longitude over the solar surface, and were placed in latitude positions consistent with the migration of the mean latitude with cycle phase. 
However, they did not consider the dependence of the mean latitude with cycle strength \citep{li03, solanki08, jiang11-a}.

The bottom panel of Fig.~\ref{TSI_1700_comp}b shows the TSI reconstruction by WLS05 based on the average of two reconstructions (S1 and S2) of the total magnetic flux using a SFTM.
The reconstruction follows the empirical model from \cite{lean00}, who used linear regressions between TSI measurements and different proxies of sunspot and facular brightness variations. Thus the sunspot darkening was derived from the yearly R$_{g}$ before 1882 and the sunspot areas after that date. The facular brightening was obtained from the photospheric flux simulated with their SFTM before 1976, and the MgII index after 1976.
In contrast, SATIRE-T2 uses one single homogeneous proxy for all magnetic features, the sunspot group number, R$_{g}$, to reconstruct the TSI over the past $\sim$300 years and semi-empirical model atmospheres to describe the brightness of each magnetic feature.

As in all SATIRE models, SATIRE-T \citep{krivova10} also employs brightness spectra computed from semi-empirical model atmospheres and weights them by the corresponding disc area coverage. The disc area coverage of spots are extracted from the daily R$_{g}$ before 1874 and sunspot group areas after 1874. To derive the filling factors of faculae and the network, the evolution of the photospheric magnetic flux is first reconstructed with the physical model of \cite{solanki00, solanki02-b} and \cite{vieira10}. This model solves a set of ordinary differential equations with the sunspot number as input to provide daily disc-integrated values of the magnetic flux in active and ephemeral regions as well as the open flux. 

To compare the TSI increase since 1700 in the three reconstructions we calculated the difference between the the year 1700 and the average over the years 1976 -- 1996.
We stopped in 1996 since the reconstruction from WLS05 ends at this time.  
The three reconstructions show a very similar increase of the TSI during this period. SATIRE-T and WLS05 show an increase of 1.3$^{+0.2}_{-0.4}$ Wm$^{-2}$ and 1.2 Wm$^{-2}$, respectively, while the increase in SATIRE-T2 (based on R$_g$) is 1.2$^{+0.2}_{-0.3}$ Wm$^{-2}$.
Both SATIRE models are in agreement during the period 1700 and 1710 because (1) between 1700 and 1710 the sun was barely active \citep[the average value of R$_{g}$ during this time was below 5;][]{hoyt98} and therefore the irradiance in both models is dominated by the quiet sun, and (2) we employed the same model atmospheres.

Figure~\ref{TSI_1700_comp} also reveals differences in the amplitude of the TSI cycles between the three reconstructions.
On the one hand, the weakest cycles (between the years 1700 and roughly 1890) have on average a slightly larger amplitude in our reconstruction than in SATIRE-T and WLS05. 
On the other hand, the amplitude of the strongest cycles (1940 -- 1960, and 1975 -- 2009) in WLS05 and our reconstructions are comparable, but lower than in SATIRE-T.  

 \begin{figure}
  \includegraphics[width=0.5\textwidth]{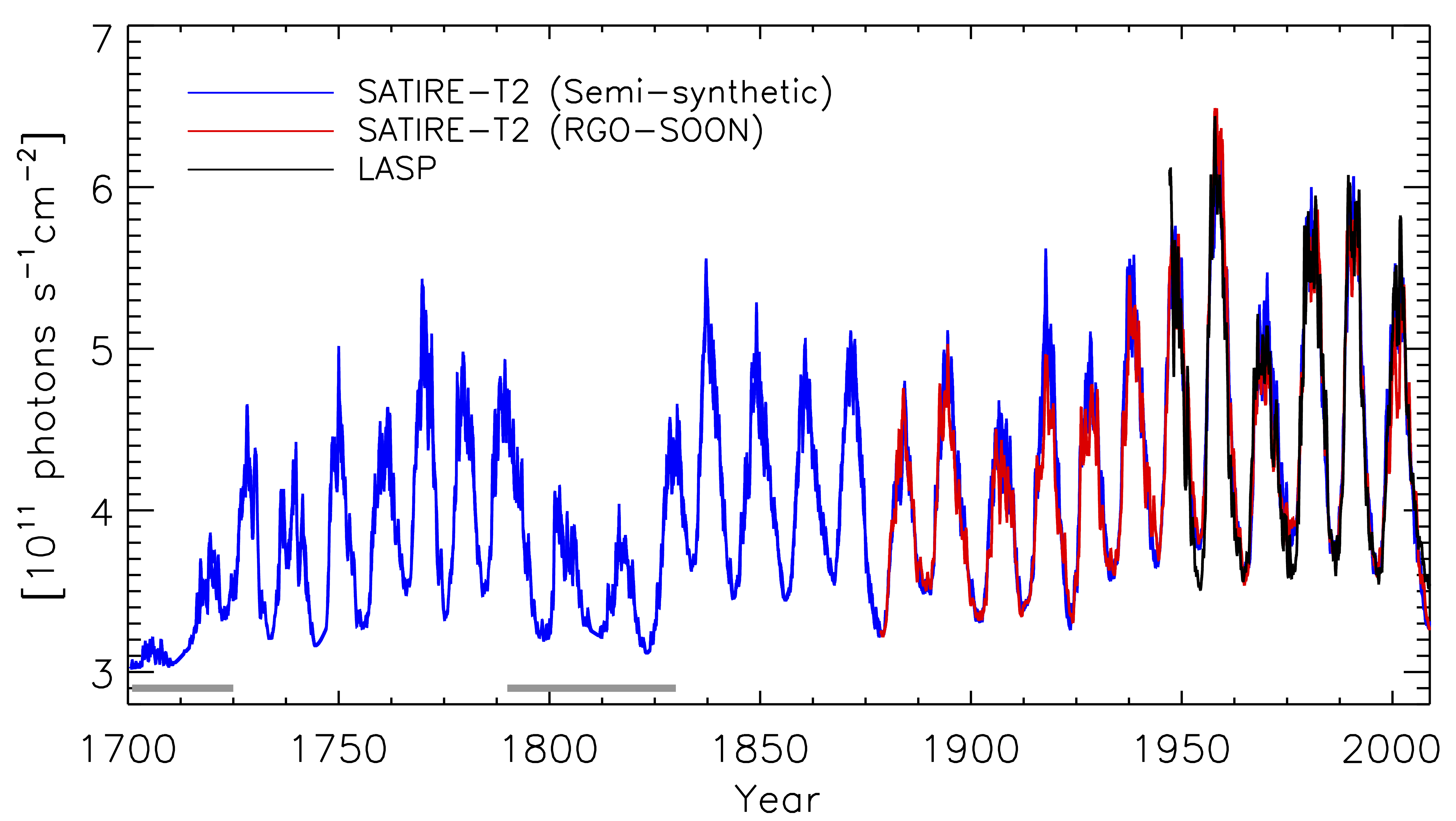}
      \caption{Lyman-$\alpha$ irradiance smoothed over 3 months. The red and blue curves correspond to the SATIRE-T2 model using as input the RGO-SOON and the semi-synthetic sunspot group records (based on the R$_g$), respectively. The grey bars at the bottom indicate the times when the statistical relationships have been extrapolated beyond the range of observations on which they are based. In black we also show the LASP Lyman-$\alpha$ composite.}
 \label{SSI_Lya}
\end{figure}

The differences could partly be due to differences in how the input data are treated, and partly due to the different photospheric magnetic flux models. 
One of the reasons for the lower amplitude of the strong cycles is the following non-linear effect present in the SFTM used here and in WLS05.
The diffusive magnetic flux transport in both SFTMs is dependent on cycle strength. During strong cycles, the number of sunspot groups on the solar surface at each time increases, and thus the groups are closer to each other, causing more flux in neighbouring active regions to cancel between each other \citep[e.g][]{baumann04}. This means that strong cycles have shorter-lived facular fields, causing both the total flux and the TSI to saturate.

Following the behaviour of the real Sun, there are also other non-linearities in the source parameters of the SFTM used here; these are a consequence of the dependence of the mean latitude and tilt angle of sunspot groups on the strength of a cycle \citep[for a comprehensive review see][]{jiang14}. 
During cycle maxima the TSI is dominated by the active region magnetic flux. 
By model construction, weak cycles have on average sunspot groups emerging at lower latitudes and with larger tilt angles. 
\cite{jiang14a} showed that active regions emerging at lower latitudes (close to the equator) and with large tilt angles cause a significant effect on the axial dipole field (and thus in the amount of net flux that is advected to the poles), so that magnetic fields live longer due to reduced cancellation between the two polarities of each active region before they are transported to the polar regions. For irradiance this means that longer-lived faculae, which raise TSI for longer spells, are generated during weak cycles.
In contrast, the magnetic flux from the ephemeral regions plays a key role in the TSI during the minima of weak cycles (see the top panel of Fig.~\ref{TSI_1700_comp}). Since weak cycles tend to have stronger following cycles \citep[which is related to the Gnevyshev-Ohl rule or Odd-Even rule;][]{gnevyshev48}, the magnetic flux from ephemeral regions after weak cycles is higher than after strong cycles due to the extended ephemeral region cycles in our model (see Sect. 2.2).

 \subsection{Spectral solar irradiance}

 \begin{figure*}
  \includegraphics[width=0.5\textwidth]{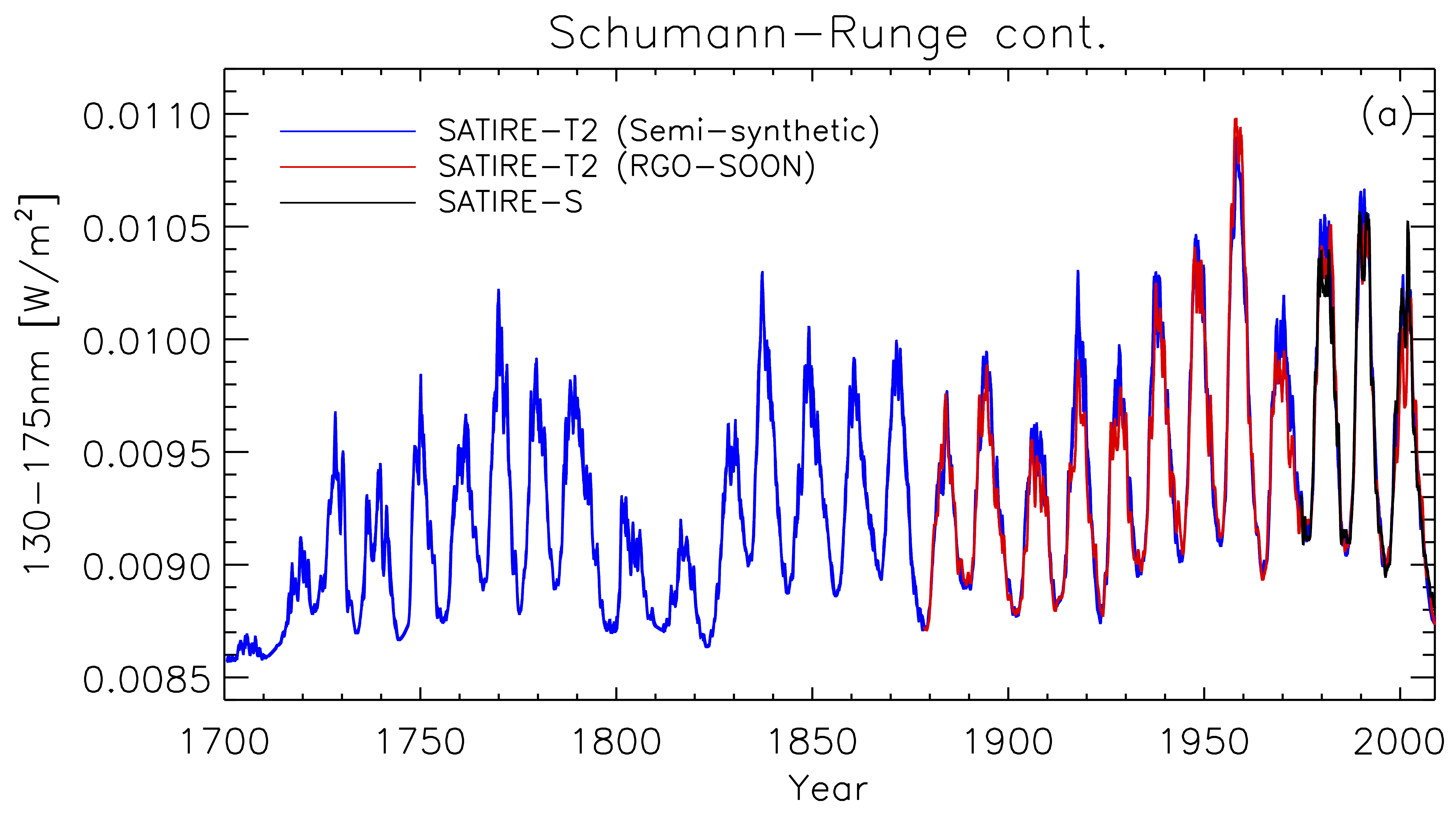}
  \includegraphics[width=0.5\textwidth]{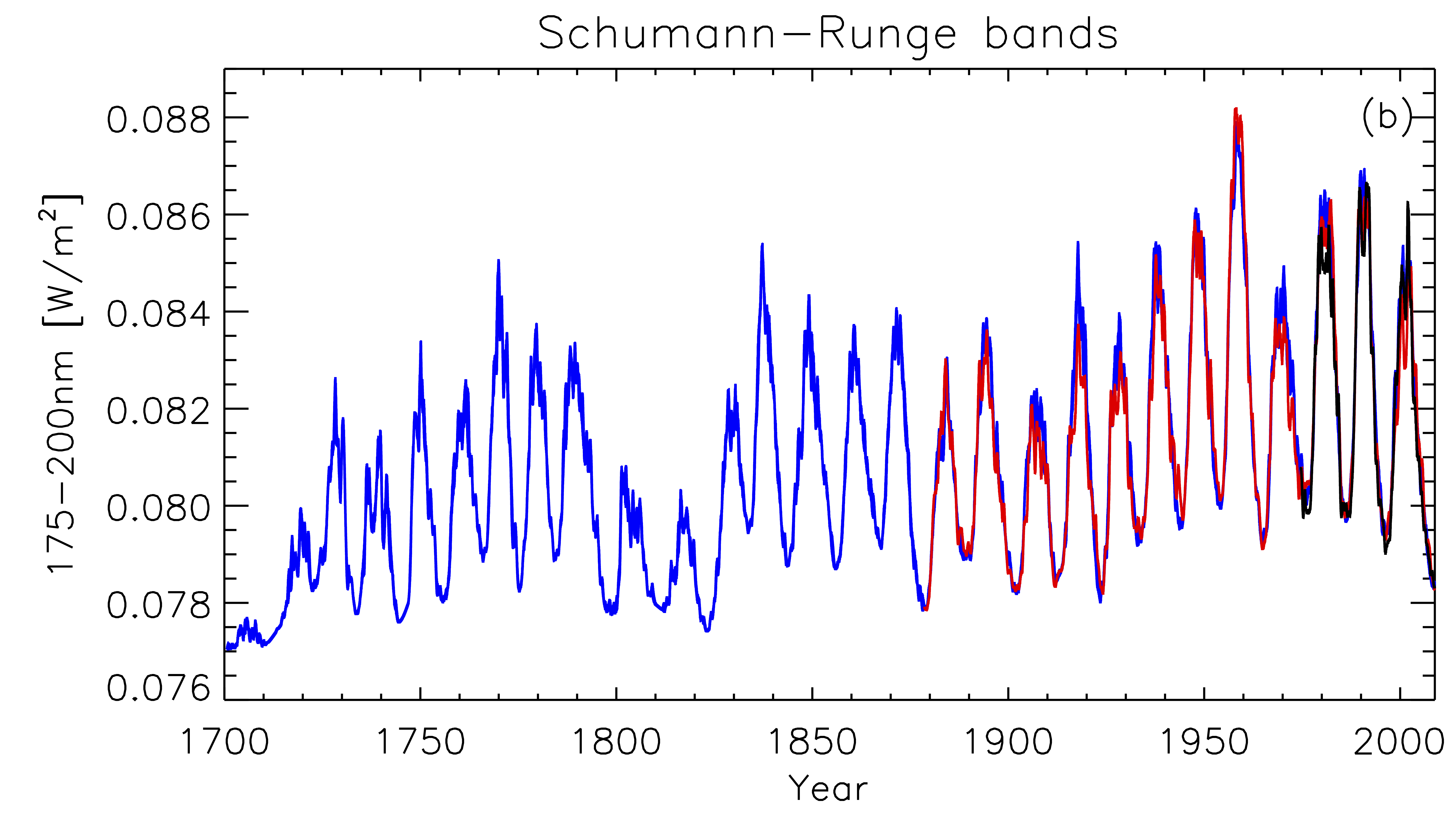}

  \includegraphics[width=0.5\textwidth]{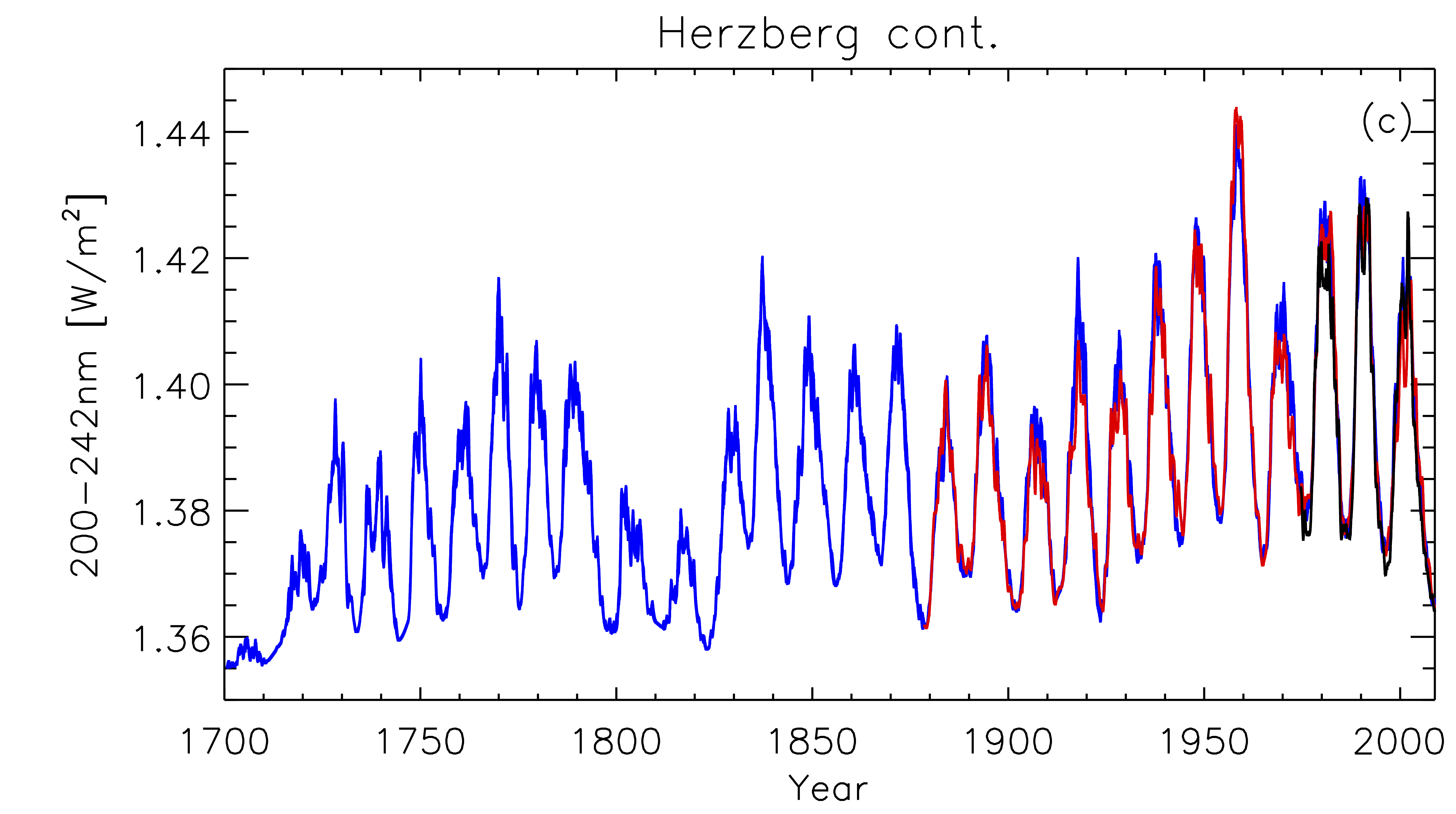}
  \includegraphics[width=0.5\textwidth]{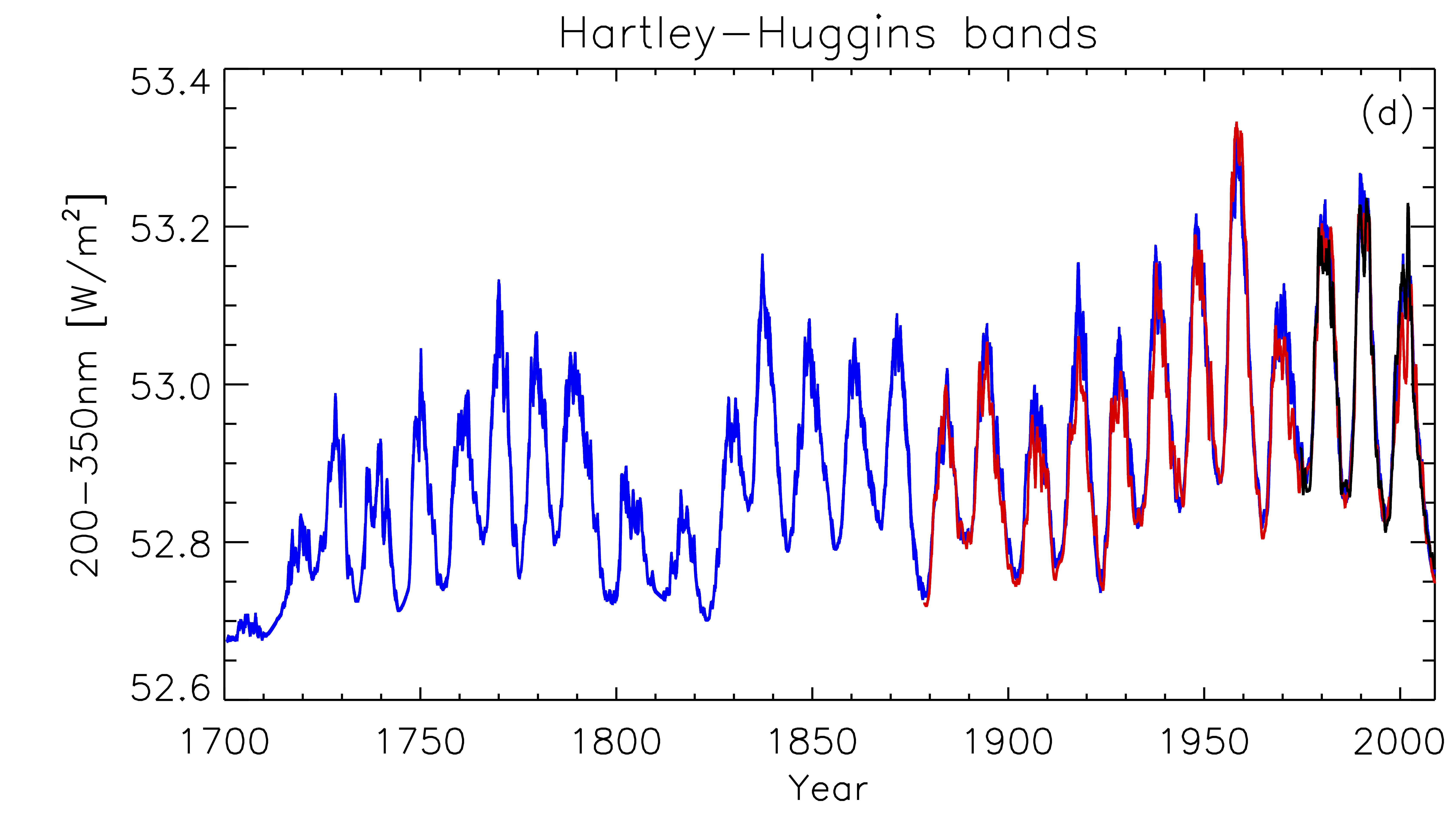}
  
  \includegraphics[width=0.5\textwidth]{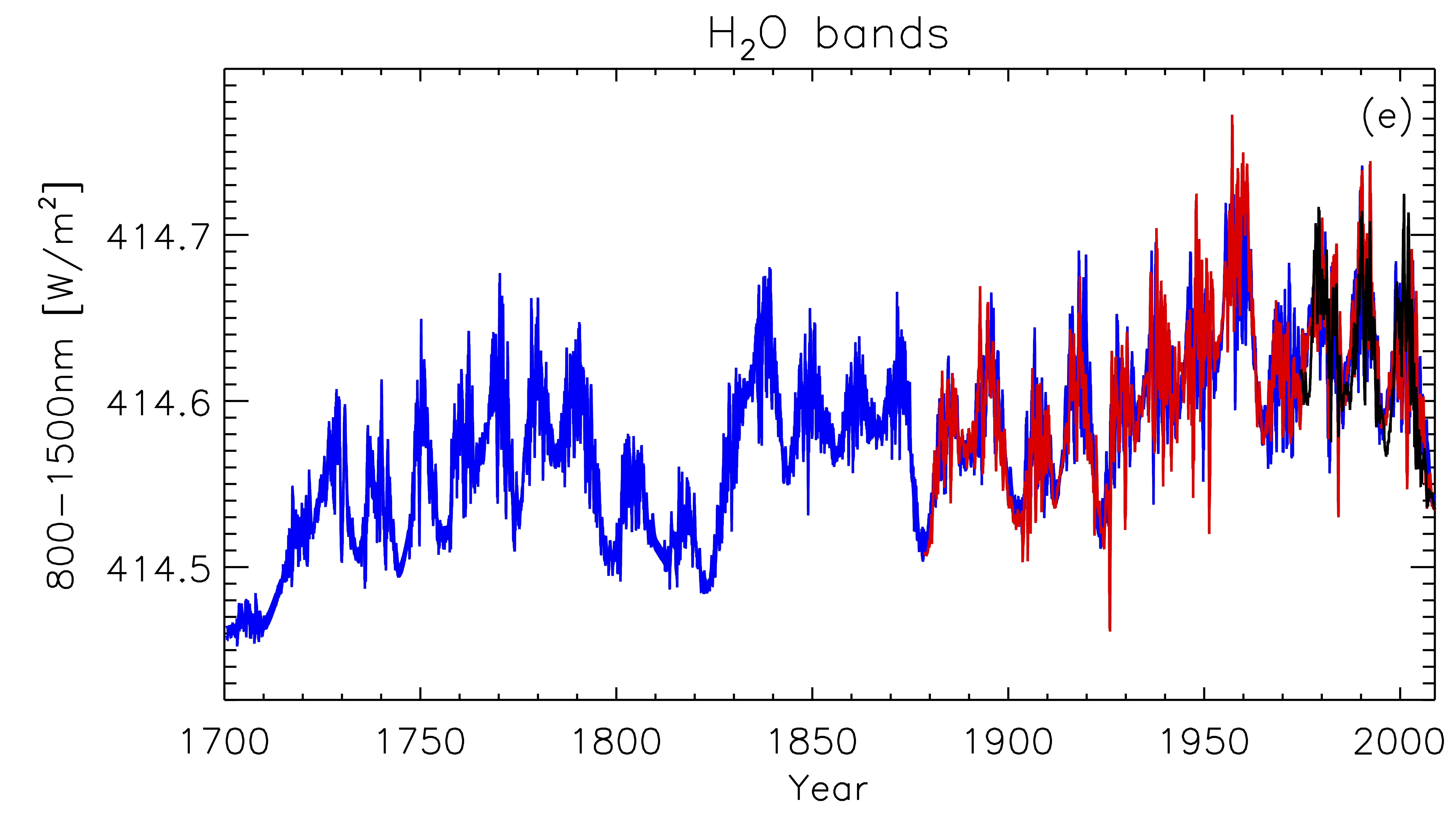}
  \includegraphics[width=0.5\textwidth]{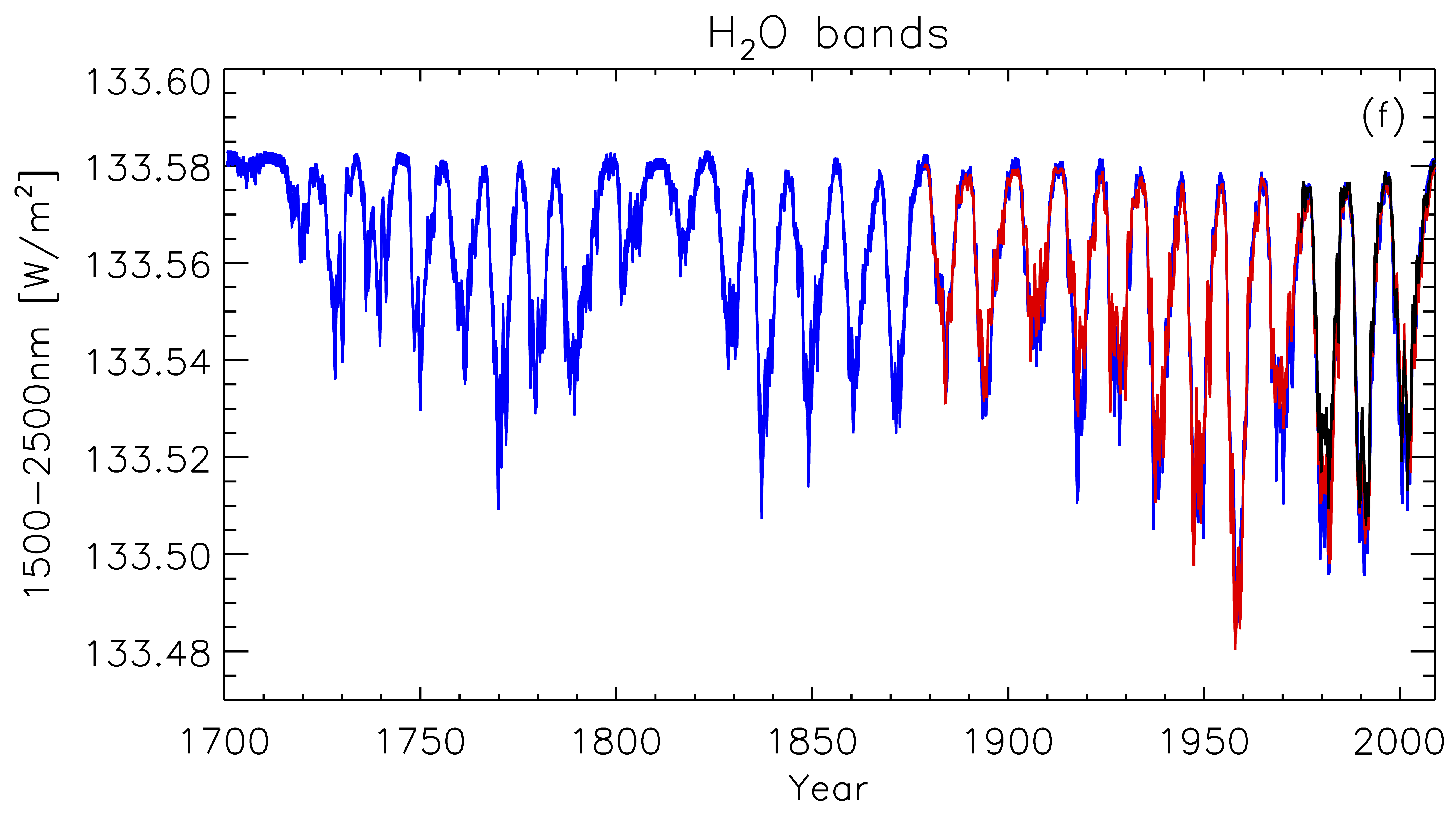}

    \caption{Reconstructed solar irradiance in various spectral bands smoothed over 1 year with SATIRE-S \citep[black;][]{yeo14}, SATIRE-T2 based on RGO data (red) and based on the synthetic data (blue): (a) Schumann-Runge oxygen continuum, (b) Schumann-Runge oxygen bands, (c) Herzberg oxygen continuum, (d) Hartley-Huggins ozone bands, (e and f) water vapour infrared bands.}
 \label{SSI_1700}
\end{figure*}
 
The semi-empirical model atmospheres used by the SATIRE family of models \citep[see Sect. 3 and][]{krivova11-b} were constructed using the ATLAS9 radiative transfer code \citep{unruh99, kurucz92}, which assumes local thermodynamical equilibrium (LTE). 

To correct for non-LTE effects in the UV \citep{krivova06, krivova09-c} we followed the approach by \cite{yeo14}.
Firstly, they offset the quiet sun UV irradiance to match the observed Whole Heliospheric Interval (WHI) reference solar spectra \citep{woods09}. Note that the variability is absolutely untouched by this measure. Secondly, below 180 nm \cite{yeo14} offset and rescaled the UV variability employing the empirical relationships based on SORCE/SOLSTICE \citep{mcclintock05, snow05} observations \citep[see][for more details]{yeo14}.

The Hydrogen Lyman-$\alpha$ solar flux (121.5 nm) is of particular interest for its impact on the photochemistry and heating in the middle atmosphere \citep[e.g.][]{WangLi13}. In Fig.~\ref{SSI_Lya} we compare the reconstructed Lyman-$\alpha$ flux (red and blue) to the LASP composite\footnote{http://lasp.colorado.edu/lisird/lya/}. The composite is based on various measurements, while proxy models were used to fill in the gaps and extrapolate the time series back to 1947 \citep{woods00}. The correlation coefficient is 0.96 between the 3-month smoothed SATIRE-T2 reconstructions and the LASP composite over the entire period of overlap. The reconstruction suggests that the Lyman-$\alpha$ irradiance has increased by about 50\% between the end of the Maunder Minimum (1700) and the average over the last three cycles.

Figure~\ref{SSI_1700} shows the reconstructed solar irradiance in various spectral bands relevant for climate studies, such as the Schumann-Runge oxygen continuum and absorption bands, the Herzberg oxygen continuum, the Hartley-Huggins ozone absorption bands, and two water vapor absorption bands.
The reconstruction employing SATIRE-S \citep{yeo14}, SATIRE-T2 based on the RGO-SOON data set, and SATIRE-T2 based on the semi-synthetic data sets are coloured, respectively, black, red, and blue.

As expected, the 11-year cycle variability, as well as the irradiance increase since the end of the Maunder Minimum, is weaker at longer wavelengths. The reversed variability with respect to the solar cycle between 1500 -- 2500 nm had already been noticed in observations and previous model results \citep[e.g.][]{krivova06, harder09, ermolli13}. This effect can be attributed to the decrease of the facular contrast at these wavelengths \citep{unruh08} such that the spot darkening exceeds the overall facular brightening. 

In Sect. 3.5 we compared the TSI and the TF using SATIRE-T2 based on RGO-SOON with these quantities obtained with SATIRE-T2 based on the semi-synthetic records. Figure~\ref{SSI_1700} demonstrates that the correspondence between the two reconstruction approaches is very good also at different wavelength ranges (note that above 180 nm the spectral irradiance is only offset and not rescaled). Since all SATIRE models use the same set of model atmospheres to represent the brightness of the magnetic features, the filling factors are the only variables that may produce a different spectral response in each of the SATIRE models. 
Therefore, any deviations between the SSI based on RGO-SOON or the semi-synthetic record, result mainly from differences in the amount of spot darkening and faculae/network brightening. For this reason differences with respect to the SATIRE-T and -S models are also expected, as both the SFTM (from which we extract the facular contribution) and the sunspot group record (from which we extract the spot darkening) are now based on statistical relationships related to the sunspot number, rather than based on direct observations.

The good match with the SATIRE-S SSI is also very encouraging, given that SATIRE-S uses the most accurate observations (magnetograms and continuum images) to compute the filling factors for each magnetic component \citep{yeo14}.
The correlation coefficient and slope of a linear fit between the daily SATIRE-S spot/facular filling factors and those of SATIRE-T2 based on RGO-SOON data (see Sect. 3) are, respectively, r = 0.97/0.92 and s = 1.05/0.82.
Note that the facular filling factors in SATIRE-S refer to all bright features, and thus include both faculae and ephemeral regions. Here we compared the filling factors of all of the bright components in each model, i.e. SATIRE-T2 facular plus ephemeral region filling factors with SATIRE-S facular filling factors.

Various studies \citep[e.g.][]{krivova06,ermolli13, thuillier14, ball14, yeo14, yeo15} have pointed out that the solar cycle variability in the UV reconstructed by SATIRE-S is much stronger than that in NRLSSI \citep{lean97, lean00}. Specifically, between 300 and 400 nm, the variation over the solar cycle in SATIRE-S is about twice that in NRLSSI \citep{yeo15}. The reconstruction presented here, being consistent with SATIRE-S, shows the same disparity to NRLSSI.

Solar cycle variability in the UV in NRLSSI is determined by matching the rotational variability in the Mg II index and the Photometric Sunspot Index, PSI, \citep{hudson82, frohlich94} to that in the UARS/SOLSTICE record \citep{rottman01} and assuming the index-to-SSI relationship so derived to apply at solar cycle timescales. In the recent review of UV SSI records and reconstructions by \cite{yeo15}, the authors noted that measured SSI rotational variability is increasingly dominated by noise towards longer wavelengths. As a direct consequence, the regression of the rotational variability in the Mg II index and PSI to observed SSI rotational variability is poor longward of 300 nm, raising doubt on the validity of the index-to-SSI relationship from such an analysis. From mimicking the NRLSSI approach on various records and reconstructions and comparing the recovered solar cycle variability to that in the original data, \cite{yeo15} demonstrated that solar cycle variability is likely underestimated in NRLSSI long-wards of 300 nm due to measurement noise in the UARS/SOLSTICE record.
Consequently, the SSI variability obtained since the end of the Maunder minimum in this study \citep[as well as that of][]{krivova10} is expected to be more reliable than the SSI variability provided by the NRLSSI.


\section{Summary}

We have reconstructed the spectral and total solar irradiance since 1700 by improving and extending the model presented by \cite{Dasi-Espuig14}. They used a SATIRE-type model together with daily magnetograms simulated with a surface flux transport model (SFTM).
The simulated magnetograms supply the evolution of the magnetic flux in active regions and their decay products. The magnetic flux that emerges within the ephemeral regions forming the network was added separately based on the concept of extended cycles, whose amplitude and length are related to those of the corresponding sunspot cycles \citep{solanki00, solanki02-b, krivova07, krivova10, vieira10}. A future improvement to SATIRE-T2 would be to include ephemeral regions in the SFTM.


In \cite{Dasi-Espuig14} the source of magnetic flux in the SFTM was based on the RGO-SOON sunspot group dataset, available since 1874. To extend the model back to 1700, we constructed semi-synthetic sunspot group records based on the sunspot number (R$_g$ and R$_z$) and statistical properties of sunspot group emergence measured by the RGO \citep{jiang11-a}.

The magnetic flux in each individual active region (sunspots and faculae) is injected into the SFTM only once - on the day of maximum area - since the SFTM describes the passive transport of the photospheric flux due to the effects of differential rotation, meridional flow, and turbulent surface diffusivity.
To calculate daily irradiances, we additionally need the position and area of the sunspot groups on the days before and after the maximum area is reached. We obtained the areas employing an empirically determined growth rate, and the observed instantaneous decay rate by \cite{hathaway08}. 

Our model has four free parameters that are fixed using observations of the TSI and the total photospheric flux during cycles 21 to 23. We tested the sensitivity of our reconstructions to uncertainties in the parameter fits as well as to the choice of a particular semi-synthetic sunspot record, and find very little change in the long-term trends of the reconstructions.

We compared our TSI reconstructions since 1700 with those from the SATIRE-T \citep{krivova10} and \citet[WLS05]{wang05}. 
We calculated the magnitude of the secular increase in TSI between the end of the Maunder Minimum and recent cycles (here taken as the average between 1976 and 1996, when all three reconstructions are available) and found that the increase is 1.2$^{+0.2}_{-0.3}$ Wm$^{-2}$, in agreement with the reconstructions from WLS05 and the SATIRE-T model. The amplitudes of the TSI cycles show some differences. In particular, the amplitude of the stronger cycles (1940 -- 1960, and 1975 -- 2009) are lower in both our reconstructions and in that of WLS05, while the amplitude of the weak cycles (1700 -- 1890) are on average slightly larger in our reconstructions. These differences are mainly due to non-linearities in the surface flux transport models used here and in WLS05 \citep[e.g.][]{jiang14}, which reflect complex processes acting in the real Sun.

Our SSI reconstructions are in good agreement with the UARS/SUSIM measurements \citep{brueckner93, floyd03-b} as well as the Lyman-$\alpha$ composite of observations \citep{woods00}.
We stress that no free parameters were adjusted to fit the UV irradiance.
We find that the spectral variability obtained with SATIRE-T2 matches that of SATIRE-S without any further parameter adjustments. This is despite their very different input data: SATIRE-S is based on detailed solar disc images and magnetograms, while SATIRE-T2 relies either on sunspot numbers (before 1878) or sunspot area records.
The good match is an example of the internal consistency between the different models of the SATIRE family. Together with their foundation in physics, this makes the SATIRE family of irradiance models particularly robust and versatile, allowing the reconstructed irradiance obtained for different timescales to be stitched together easily to obtain a seamless and internally consistent record of SSI and TSI over very long periods of time, with the quality determined only by the quality of the underlying data available for a particular period.
In particular, according to the analysis of \cite{yeo15}, the SSI provided by the SATIRE family of models should be more reliable than that obtained by NRLSSI \citep{lean97, lean00}. 

The complete total and spectral (115 nm -- 160 $\mu$m) irradiance reconstructions since 1700 will be available from http://www2.mps.mpg.de/projects/sun-climate/data.html.

\begin{acknowledgements}
We thank Y-M. Wang for providing their total solar irradiance reconstruction and Robert Cameron for helpful discussions. The research leading to these results has received funding from the European Community's Seventh Framework Programme (FP7 2012) under grant agreement number 313188 (SOLID). J.J acknowledges financial support from the National Natural Science Foundations of China through grant 11173033 and 11522325, and Y.U. from grant ST/K001051/1 from the Science and
Technology Facilities Council. This work was partly supported by the BK21 plus program through the National Research Foundation (NRF) funded by the Ministry of Education of Korea. We also thank the referee whose comments helped to make the manuscript clearer.
\end{acknowledgements}

 \bibliographystyle{aa.bst}
 \bibliography{marybib.bib}

\end{document}